\documentclass[a4paper,fleqn,usenatbib]{mnras}
\usepackage{newtxtext,newtxmath}
\usepackage[T1]{fontenc}
\usepackage{ae,aecompl}

\usepackage{graphicx}	
\usepackage{amsmath}	
\usepackage{amssymb}	
\usepackage{verbatim,graphicx,epsfig,dcolumn,xcolor}

\newcolumntype{.}{D{.}{.}{4}}
\newcolumntype{,}{D{.}{.}{2}}
\newcolumntype{;}{D{.}{.}{1}}

\newcommand{\appropto}{\mathrel{\vcenter{
  \offinterlineskip\halign{\hfil$##$\cr
    \propto\cr\noalign{\kern2pt}\sim\cr\noalign{\kern-2pt}}}}}

\title[A dusty wind from the giant RU Vul]{Circumstellar CO $J = 3 \rightarrow 2$ detected around the evolving metal-poor ([Fe/H] $\approx$ --1.15 dex) AGB star RU Vulpeculae}
\author[I. McDonald et al.]{I.~McDonald$^{1}$\thanks{E-mail: mcdonald@jb.man.ac.uk}, S.~Uttenthaler$^{2}$, A.~A.~Zijlstra$^{1,3}$, A.~M.~S.~Richards$^1$, E.~Lagadec$^4$\\
$^{1}$Jodrell Bank Centre for Astrophysics, Alan Turing Building, Manchester, M13 9PL, UK\\
$^{2}$Kuffner Observatory, Johann-Staud-Stra{\ss}e 10, 1160 Wien, Austria\\
$^{3}$Department of Physics \& Laboratory for Space Research, University of Hong Kong, Pokfulam Road, Hong Kong\\
$^{4}$Laboratoire Lagrange, Universit\'e C\^ote d'Azur, Observatoire de la C\^ote d'Azur, CNRS, Boulevard de l'Observatoire, CS 34229, 06304 Nice Cedex 4, France\\
}

\date{Accepted XXX. Received YYY; in original form ZZZ}

\pubyear{2019}

\begin{document}
\label{firstpage}
\pagerange{\pageref{firstpage}--\pageref{lastpage}}
\maketitle

\begin{abstract}
We report the first detection of CO $J = 3 \rightarrow 2$ around a truly metal-poor evolved star. RU Vulpeculae is modelled to have $T_{\rm eff} \approx 3620$ K, $L \approx 3128 \pm 516$ L$_\odot$, log($g$) = 0.0 $\pm$ 0.2 dex and [Fe/H] = --1.3 to --1.0 dex, and is modelled to have recently undergone a thermal pulse. Its infrared flux has approximately doubled over 35 years. ALMA observations show the 3$\rightarrow$2 line is narrow (half-width $\sim$1.8--3.5 km s$^{-1}$). The 2$\rightarrow$1 line is much weaker: it is not confidently detected. Spectral-energy-distribution fitting indicates very little circumstellar absorption, despite its substantial mid-infrared emission. A VISIR mid-infrared spectrum shows features typical of previously observed metal-poor stars, dominated by a substantial infrared excess but with weak silicate and (possibly) Al$_2$O$_3$ emission. A lack of resolved emission, combined with weak 2$\rightarrow$1 emission, indicates the dense circumstellar material is truncated at large radii. We suggest that rapid dust condensation is occurring, but with an aspherical geometry (e.g., a disc or clumps) that does not obscure the star. We compare with T UMi, a similar star which is currently losing its dust.
\end{abstract}




\section{Introduction}
\label{IntroSect}

\subsection{Mass loss from metal-poor stars}
\label{IntroMdotSect}

The vast majority of stars undergo a terminal dust-laden wind on the asymptotic giant branch (AGB; \citealt{HO18}). Canonically, winds from single stars are driven by three primary mechanisms: (1) magneto-acoustic heating of a warm chromosphere above the stellar surface \citep[e.g.][]{DHA84}; (2) levitation of surface material by pulsations, which can then go on to form dust and (3) be radiatively accelerated from the star, either by absorption \citep[e.g.][]{Willson00}, or by scattering if the grains are large enough \citep{Hoefner08}. Collisional coupling between dust and gas ensures both media are ejected from the star.

Mass-loss mechanisms and prescriptions invoking only magneto-acoustic heating \citep{Reimers75,SC05,CS11} fail to reproduce the mass-loss rates of pulsating stars \citep[e.g.][]{DBDdK+10}, or the radial acceleration profiles of their outflows \citep{DJDB+10}. Pulsational piston velocities are $\sim$10 km s$^{-1}$ \citep{Hinkle78,HHR82,HLS97,LHK00,LWH+05}, and fall short of the $\sim$30--60 km s$^{-1}$ escape velocities of these stars. Hence, it is assumed that radiation pressure on dust dominates driving in all dust-producing AGB stars. However, many stars have insufficient absorption for this to work \citep{Woitke06b}, and models have only reproduced dust-driven winds around the most extreme stars \citep[e.g.][]{BHAE15}. Scattering may increase the computed dust absorption \citep{Hoefner08,NTI+12}. However, the general expectation is that it is more difficult to grow grains this large around less extreme (lower mass-loss rate, luminosity or metallicity) stars, due to the presumed reduction in the frequency of collisions between refractory particles \citep[cf.][]{DAGHS+17}.

Metal-poor stars have less atmospheric opacity, so are smaller and hotter \citep[e.g.][]{MGB+08}, while the dust-condensation radius should change very little\footnote{If the condensation temperature, $T_{\rm cond}$, does not appreciably change, then the condensation radius should not either, as $R_{\rm cond}^2  \appropto L T_{\rm cond}^{-4}$. Note that there will still be some changes due to the wavelength-dependent absorption of light \citep{BH12}.}. This decreases the strength of pulsations, and increases the gravitational barrier that material must overcome before it reaches the dust-condensation zone. Simultaneously, the lack of refractory metals decreases the dust:gas ratio, making it much harder for dust to drive a wind. In carbon stars, the carbon enhancement provided by third dredge-up allows amorphous carbon dust to drive the wind: indeed carbon enhancement may trigger the superwind\footnote{Following \citep{RV81}, this is defined as a wind with a mass-loss rate in substantial excess of the \citet{Reimers75} law.} at low metallicities \citep{LZ08,NBMG13}. However, the mass-loss mechanism of metal-poor, oxygen-rich stars is unclear. Understanding this is an important issue as, due to hot bottom burning \citep[e.g.][]{KL14}, oxygen-rich stars will be the first AGB stars to produce dust in the early Universe, where large amounts of dust are seen in young galaxies \citep[e.g.][]{BCC+03,BCB+06,MMH+10,CCJ+15}. The contribution of AGB stars to this dust is poorly determined.

Despite the supposed difficulties in producing dust, metal-poor stars are prodigious dust producers. Dust is observed around luminous, metal-poor stars in both nearby dwarf galaxies \citep[e.g.][]{MZBS+07,BSvL+09,BMB+15,SML+12,MZS+13,JMB+18,GBM+19b} and many globular clusters \citep{vLMO+06,LPH+06,MvLD+09,BMvL+09,BvLM+10,SMM+10,MBvL+11,MvLS+11}. So far, there has been no clear observational evidence that dust production by oxygen-rich stars\footnote{It is both expected and observed that \emph{carbon-}rich stars produce similar amounts of dust at all metallicities \citep[e.g.][]{SML+12,BMS+15,JMB+18,BEM+19,GBM+19}. This is because the (third) convective dredge-up process that occurs during thermal pulses brings carbon to the stellar surface, thus the stars generate their own refractory materials, regardless of their initial metallicity.} is any less efficient at low metallicities \citep{MBG+19,GBM+19} nor, conversely, any more efficient at super-solar metallicities \citep{vLBM08}.

This suggests that mass-loss rates of many oxygen-rich stars are determined by the star's pulsations \citep{MZ16,MBG+19,MT19}, rather than the effectiveness of radiation pressure on dust. However, mid-infrared spectra of globular cluster stars suggest that their dust has a very different composition to solar-metallicity stars \citep{SMM+10,MvLS+11,JKS+12}, suggesting the dust around metal-poor stars can achieve a higher opacity \citep{MBvLZ11,MBG+19}. Consequently, with only mid-infrared observations, it is not clear whether the mass-loss rate of oxygen-rich stars is truly set by pulsations (and thus independent of metallicity), or whether this correlation is an artefact of using dust-column density as a proxy for the star's total mass-loss rate.

\subsection{Previous carbon monoxide observations}

The optical properties, condensation fraction (dust:gas ratio) and outflow velocity of dust from metal-poor stars suffer from a severe lack of empirical data \citep[e.g.][]{MBvLZ11}. These properties respectively set the dust optical depth, its conversion to a mass column density, and the subsequent conversion to a mass-loss rate. Subject to second-order uncertainties, the wind properties can be better calibrated if a gas mass-loss rate and expansion velocity are observed from the intensities and widths of millimetre-range CO lines \citep[e.g.][]{MdBZL18}. Observations of mildly metal-poor stars in the Large Magellanic Cloud (LMC; [Fe/H] $\approx$ --0.3 dex) hint at a declining gas outflow velocity with decreasing metallicity, but these observations are restricted to carbon stars \citep{GVM+16,MSS+16} and indirect measurements from OH masers around the most-luminous, most-evolved AGB stars \citep{GvLZ+17}.

CO observations towards low-mass, metal-poor stars are mostly limited to globular clusters, where the CO appears to be dissociated by the strong interstellar UV field \citep{MZ15a,MZL+15}. The one successful CO observation (47 Tuc V3; [Fe/H] = --0.7 dex) suggests the outflow velocity is slower than would be expected from a Galactic disc star, hinting that the outflow velocity remains set by radiation pressure on dust \citep{MBG+19}, and corroborating the aforementioned measurement in the LMC. If metal-poor stars exhibit a similar mass-loss rate to otherwise-identical metal-rich stars (Section \ref{IntroMdotSect}), but their winds remain dust driven, then their decreased dust:gas ratios will mean less momentum is transferred from stellar radiation to the dusty wind, leading to a decrease in the wind's velocity. However, it is not clear whether an outflow can become arbitrarily slow, nor how high the opacity of dust can become, before the mass-loss rate is forced to decline due to infall of stagnant material back onto the star.

\subsection{Identifying new targets}

More extreme environments can be tested by observing stars at even lower metallicities. The search for truly metal-poor stars\footnote{Following \citet{BC05}, we define a metal-poor star as having [Fe/H] $<$ --1.0 dex.} now leads us to the Galactic halo, which is close enough that CO observations can be made. Among the literature observations of possible halo giants, one stands out: RU Vulpeculae. \citet{UGT16} show RU Vul to be an oxygen-rich giant with a metallicity of [Fe/H] $\approx$ --1.6 dex, and identify that it is undergoing the initial phases of a thermal pulse. RU Vul has a very substantial infrared excess ($K_{\rm s}-[22] = 3.295$ mag; \citealt{CWC+13}), making it unusually dusty compared to Galactic (solar-metallicity) stars ($K_{\rm s}-[22] \sim 2$ mag for the bulk of stars, i.e.\ those with pulsation periods of $60 \lesssim P \lesssim 300$ days; \citealt{MZ16}), and placing it among the most-metal-poor dust-producing giant stars known. Consequently, we adopted it as an excellent target with which to understand metal-poor stellar winds.

In this work, we will explore the properties of RU Vul and its wind, and try to place it in the context of observations of other metal-poor stars. The remainder of this paper is summarised as follows:
\begin{itemize}
\item Section \ref{SpecSect} re-analyses literature data on RU Vul to obtain accurate stellar parameters;
\item Section \ref{ALMASect} presents new observations of RU Vul with the Atacama Large Millimetre/sub-millimetre Array (ALMA) and discusses the star's gaseous wind properties;
\item Section \ref{DustSect} discusses the ALMA continuum data and RU Vul's dust production;
\item Section \ref{DiscSect} discusses our findings and presents our interpretations of the stellar system; and
\item Section \ref{ConcSect} concludes the paper.
\end{itemize}


\section{Stellar properties}
\label{SpecSect}

\subsection{Background}
\label{PropBackSect}

\begin{figure}
\centerline{\includegraphics[height=0.45\textwidth,angle=-90]{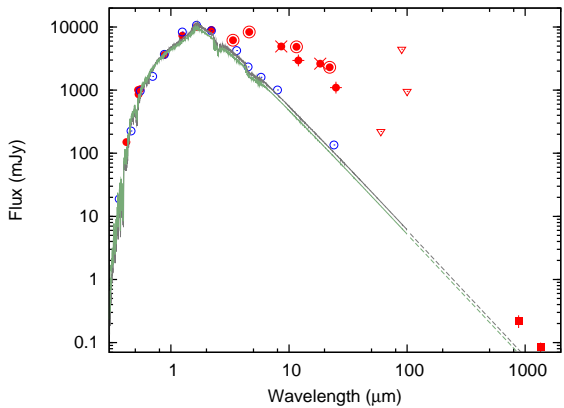}}
\caption{Spectral energy distribution of RU Vul (red, solid points), corrected for interstellar reddening. A comparison star, $\omega$ Cen LEID 42044 (blue, hollow points) is shown, multiplied by a factor of 16. Overplotted as lines are two {\sc bt-settl} model spectra \citep{AGL+03} at 3600 K (top, grey) and 3700 K (bottom, green), both with log($g$) = 0 dex and [Fe/H] = --1.5 dex, representing the spectroscopically derived properties of the stellar photosphere (Section \ref{SpecAtmosSect}). Short-wavelength data adheres to this photosphere, indicating no optical absorption; long-wavelength data exceeds the photospheric flux, showing emission from reprocessing of radiation by circumstellar dust. Infrared data on RU Vul has additional symbols: $+$ = \emph{IRAS}, $\times$ = \emph{Akari}, $\odot$ = \emph{WISE}, showing the increase in infrared flux over time (see also Table \ref{IRTable}); ALMA continuum measurements from this work are shown as squares; triangles show upper limits. Optical data are sourced from \emph{Hipparcos} \citep{vanLeeuwen07}, \emph{Tycho} \citep{Perryman97}, The Amateur Sky Survey (TASS) Mark IV catalogue \citep{DRSC06}, and the Two Micron All Sky Survey (2MASS; \citealt{SCS+06}).}
\label{SEDFig}
\end{figure}

\begin{center}
\begin{table*}
\caption{Mid-infrared observations of RU Vul showing its increase in infrared emission over time and, for comparison, the decrease in emission from T UMi.}
\label{IRTable}
\begin{tabular}{lcccccccccc}
    \hline \hline
    \multicolumn{1}{c}{Satellite/} & \multicolumn{1}{c}{Epoch} & \multicolumn{1}{c}{$\lambda_{\rm eff}$} & \multicolumn{2}{c}{RU Vul} & \multicolumn{2}{c}{T UMi} \\
    \multicolumn{1}{c}{Filter}     & \multicolumn{1}{c}{(yr)}  & \multicolumn{1}{c}{($\mu$m)}            & \multicolumn{1}{c}{$F_\nu$}  & \multicolumn{1}{c}{$F_\nu \lambda^2$} & \multicolumn{1}{c}{$F_\nu$}  & \multicolumn{1}{c}{$F_\nu \lambda^2$} \\
    \multicolumn{1}{c}{\ }         & \multicolumn{1}{c}{\ }    & \multicolumn{1}{c}{\ }                  & \multicolumn{1}{c}{(Jy)}     & \multicolumn{1}{c}{(Jy $\mu$m$^2$)} & \multicolumn{1}{c}{(Jy)}     & \multicolumn{1}{c}{(Jy $\mu$m$^2$)} \\
    \hline
 \emph{IRAS} [12]  & 1983 & 10.15            & 2.93 $\pm$ 0.12   & 302 $\pm$ 12     & 14.4 $\pm$ 0.597  & 1483 $\pm$ 62 \\ 
 \emph{IRAS} [25]  & 1983 & 21.73            & 1.09 $\pm$ 0.06   & 515 $\pm$ 28     & 5.19 $\pm$ 0.311  & 2450 $\pm$ 147 \\ 
 \emph{IRAS} [60]  & 1983 & 51.99            & 0.22 $\pm$ 0.05$^1$& 595 $\pm$ 135   & 0.762 $\pm$ 0.046 & 2060 $\pm$ 124 \\
 \emph{IRAS} [100] & 1983 & 95.30            & $\lesssim$0.96    & $\lesssim$8700   & 0.516 $\pm$ 0.103 & 4686 $\pm$ 935 \\
 \emph{DIRBE} [2.2]& 1991 & 2.20             & \                 & \                & 36.7 $\pm$ 4.9    & 178 $\pm$ 24 \\ 
 2MASS             & 1999 & 2.16             & 8.76 $\pm$ 0.16   & 40.8 $\pm$ 0.7   & 46.4 $\pm$ 11.8   & 216 $\pm$ 55 \\ 
 \emph{Akari} [9]  & 2007 & 8.23             & 4.847 $\pm$ 0.031 & 328 $\pm$ 2      & 6.318 $\pm$ 0.0347& 428 $\pm$ 2 \\
 \emph{Akari} [18] & 2007 & 17.61            & 2.605 $\pm$ 0.052 & 808 $\pm$ 16     & 2.732 $\pm$ 0.045 & 847 $\pm$ 14 \\
 \emph{Akari} [90] & 2007 & 76.90            & $\lesssim$4.5    & $\lesssim$36\,000 & $\lesssim$4.5     & $\lesssim$36\,000 \\
 \emph{WISE} [11]  & 2010 & 10.79            & 4.812 $\pm$ 0.047 & 560 $\pm$ 5      & 3.925 $\pm$ 0.051 & 457 $\pm$ 6 \\ 
 \emph{WISE} [22]  & 2010 & 21.92            & 2.284 $\pm$ 0.044 & 1097 $\pm$ 21    & 1.784 $\pm$ 0.029 & 857 $\pm$ 14 \\ 
\hline
\multicolumn{5}{p{0.47\textwidth}}{$^1$Since this is a $<$5$\sigma$ detection, it is shown as an upper limit in Figure \ref{SEDFig}.}\\
\hline
\end{tabular}
\end{table*}
\end{center}

\subsubsection{Discovery, infrared brightening and thermal pulses}
\label{PropBackDiscSect}

RU Vul was first identified as a variable by \citet{Wolf1904}, but the first documentation of its pulsation was by \citet{Beyer28}, who noted a period of 158.3 days with a photographic amplitude of $\sim$1.8 mag. Data from the American Association of Variable Star Observers (AAVSO)\footnote{\tt http://www.aavso.org} from the 1930s onwards shows periodic variability between visual magnitudes of $\sim$9.0 and $\sim$11.5. The period begins to decline around 1955 towards today's value of $\sim$108 days \citep{RSM+12,UGT16}. A marked brightening of photometric minimum to 10$^{th}$ magnitude occurred around 1965 (the maximum did not change substantially), upon which the pulsation amplitude decreased to today's $V$-band semi-amplitude of $\sim$0.39 mag. The amplitude is now close to the visual scatter recorded by the AAVSO, and insufficient data exists to extract an updated period from \emph{Gaia} Data Release 2 \citep{GaiaDR2}.

There is insufficient data at infrared wavelengths to determine how the bolometric luminosity of the star is changing. The star is (on average) clearly becoming optically brighter and less variable, indicating an apparent rise in temperature\footnote{Optical variability in late-type, oxygen-rich stars is dominated by molecular opacity effects, particularly of TiO \citep[e.g.][]{BHAE15}. The opacity of the TiO bands has a highly non-linear behaviour with temperature. In cooler stars, a given radial pulsation will produce a significantly larger optical variability due to this molecular blanketting. Therefore, a warming star will exhibit both an increase in optical flux and a decreasing visual pulsation amplitude.} and decrease in radius. Changes in pulsation period reflect changes in the sound travel time in the stellar atmosphere. These come from a combination of changes in the stellar structure following the thermal pulse, and changes to the stellar radius itself. Taking at face value the relation of \citep{Wood90}, $\log P \propto 1.94 \log R$, a period decline of $\Delta \log P = 0.166$ dex potentially translates to a decrease in radius of 18 per cent since 1955.

Simultaneously, the infrared flux seems to be increasing (Figure \ref{SEDFig}). Table \ref{IRTable} lists the space-based infrared observations of RU Vul, from the {\it Infrared Astronomical Satellite} ({\it IRAS}; \citealt{BHW88}), {\it Akari}  \citep{KAC+10}, and the \emph{Wide-Field Survey Explorer} (\emph{WISE}; \citealt{CWC+13}). These include a measure in Rayleigh--Jeans units ($F_\nu \lambda^2 \propto F_\nu \nu^{-2}$). This measure has some wavelength dependency due to the finite temperature of dust, but much less so than simply using $F_\nu$. The comparison of (e.g.) \emph{IRAS} [12] versus \emph{WISE} [11] and \emph{IRAS} [25] versus \emph{WISE} [22] show that $F_\nu \lambda^2$ approximately doubles over the 27-year observing window: $F_{2010} / F_{1983} \approx 1.85$ at $\sim$12 $\mu$m and $\approx$2.13 at $\sim$25 $\mu$m.

There are two likely explanations for the infrared brightening: either (1) because the star has optically brightened, pre-existing dust is reprocessing more optical light into the infrared, or (2) more dust has formed. If existing material has warmed, the increase in flux at short wavelengths should be greater than at longer wavelengths. However, Table \ref{IRTable} shows the reverse to be true: the increase in $F_\nu \lambda^2$ at 22--25 $\mu$m is substantially larger than at 11--12 $\mu$m. Changes in the spectral slope between 11- and 25 $\mu$m could also come from changes in dust properties (e.g., grain mineralogy, size or porosity). However, without condensing new dust, we consider this unlikely. Hence, we conclude that the increase in infrared flux represents significant and rapid dust condensation around RU Vul. The dust emissivity has doubled over the 27 years since 1983 and, if it has cooled, the dust volume should be increased by even more. In the simple approximation that the 25-$\mu$m flux has been linearly increasing due to constant (optically thin) dust formation, we can extrapolate the onset of rapid dust formation to within a few years of 1956 (cf., the start of the period decline around 1955) and that, prior to this, RU Vul may have been an unremarkable star in terms of its dust properties (this concept is discussed further in Sections \ref{DiscWindSect} and \ref{DiscPChangeSect}).

\citet{UGT16} interpret the period decline as the start of a thermal pulse cycle. Here, runaway helium burning shuts off hydrogen burning, causing the star to initially shrink, until the energy from the detonation reaches the stellar surface. They derive a metallicity of [Fe/H] = --1.59 $\pm$ 0.05 dex for RU Vul, based on a small region of the spectrum, which could potentially be affected by the star's short- and long-term out-of-equilibrium state. Equally, their luminosity of 2830 $\pm$ 520 L$_\odot$ and consequent distance derivation of 2070 $\pm$ 130 pc is based on comparison to a period--$K_{\rm s}$-band brightness relationship. Period and brightness were not measured simultaneously, which is problematic when both appear to be changing \citep{UvSV+11}.

\subsubsection{The distance to RU Vul}
\label{PropBackDistSect}

Determining the parallax of variable, red stars is compounded by problems of correctly weighting measurements and accounting for changes in the centre of light of the star's disc which, being $\sim$1 AU in radius, has a similar angular size to the parallax effects being measured. The treatment of AGB stars in \emph{Gaia} DR2 is not yet perfect \citep[e.g.][]{MDBLZ18}, meriting a detailed treatment of RU Vul and separation of the available \emph{Hipparcos} and \emph{Gaia} data. The distance to RU Vul is unconstrained by the \emph{Hipparcos} parallax of $\varpi = 0.93 \pm 1.70$ mas. The \emph{Gaia} Data Release 1 (DR1) parallax of $\varpi = 0.30 \pm 0.51$ mas includes the \emph{Hipparcos} data \citep{GaiaDR1}, and effectively limits the distance to over $\sim$1 kpc but still does not provide an upper limit to the star's distance. \emph{Gaia} Data Release 2 is based solely on \emph{Gaia} data \citep{GaiaDR2}, and provides $\varpi = 0.540 \pm 0.059$ mas. The quality of the DR2 fit is good ( thus we can identify no reason not to adopt it), and the fractional uncertainty (11 per cent) is relatively small (meaning we need not be concerned by significant probability of negative parallaxes). We can therefore simply invert the parallax to provide $d = \varpi^{-1} = 1851^{+204}_{-184}$ pc.

Despite potential issues with both methods, the distances from the pre-decline pulsation period ($2070 \pm 130$ pc; Section \ref{PropBackDiscSect}; \citealt{UGT16}) and parallax (\emph{Gaia}) agree within uncertainties, confirming RU Vul as a fundamental-mode pulsator (sequence $C$ in, e.g., \citet{Wood15}). In the following, we adopt a fixed distance of $d = 2000 \pm 165$ pc, based on the combination of pulsation and parallax data, which offer two independent distance estimates.

Orbital integration by \citet{MB18} indicate that the star progresses on an elliptical orbit around the Galaxy ($e \approx 0.4$), confined to within $|z| \approx 1.5$ kpc of the Galactic Plane, making the star likely a thick disc star, rather than a halo star (cf.\ \citealt{UGT16}). An origin in the thick disc implies an age of $\sim$9--12 Gyr \citep[e.g.][]{KMH+17}, hence an initial mass of $\sim$0.78--0.92 M$_\odot$, depending on its exact metallicity \citep{MBG+19}.

Obtaining accurate stellar parameters for out-of-equilibrium stars like RU Vul is challenging, as direct, simultaneous measurements are difficult to obtain. In the remainder of this section, we go through different ways of constraining the star's parameters in an attempt to improve them.

\subsection{Atmospheric parameters from spectra}
\label{SpecAtmosSect}

\begin{figure*}
\centerline{\includegraphics[height=0.95\textwidth,angle=-90]{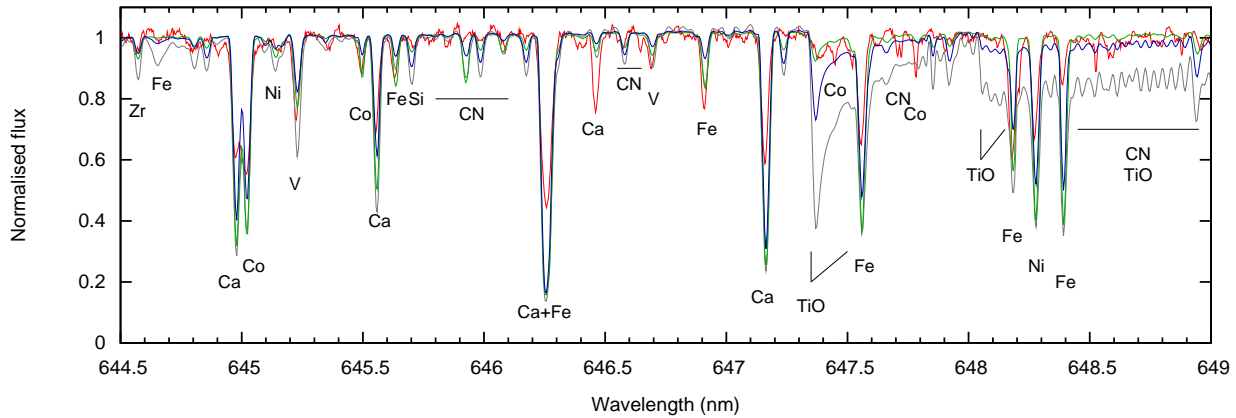}}
\caption{Spectrum of RU Vul from \citep{UvSV+11} (red line), showing a close-up of the TiO bandhead. Overplotted in grey is a {\sc bt-settl} model spectrum at 3600 K, log($g$) = 0 dex and [Fe/H] = --1.5 dex. The green spectrum is identical, at 3800 K; and the blue spectrum is identical, at [Fe/H] = --2 dex.}
\label{SpecFig}
\end{figure*}

\subsubsection{Spectral synthesis}
\label{SpecAtmosSynthSect}

We first re-analyse the optical spectrum published by \citet{UvSV+11}, performing a spectrum synthesis and abundance analysis. This followed the methodology of \citet{JP10}, which is outlined in the next paragraph. We also used their line lists, with the adjustments to log($gf$) included in \citet{JMP+15} and \citet{JRP+15}. Lines were restricted to $\lambda > 6000$ \AA, as shorter wavelengths have much lower signal, causing scatter in the derived parameters.

While the TiO lines in the spectrum are generally weak, they can still affect the weaker metal lines, which are needed to constrain the micro-turbulent velocity, $v_{\rm t}$. Consequently, the quality of the fit is relatively low compared to the spectrum's signal-to-noise. Out-of-local-thermodynamic-equilibrium (non-LTE) effects were seen in the spectrum, including line doubling and emission in chromospherically active lines, which may cause uncharacterised errors in the final fit. No evidence of products of third dredge-up were visible (e.g., the ZrO band expected at $\sim$646.75 nm in Figure \ref{SpecFig} or a strong Li 6707 line).

A standard equivalent-width analysis of Fe {\sc i} and Fe {\sc ii} lines was performed, and Gaussian profiles were fit to lines present in the continuum-normalised spectra. Blended lines were fit with multiple Gaussian components. The resulting equivalent widths were analysed using the {\tt abfind} task in {\sc moog}\footnote{http://www.as.utexas.edu/~chris/moog.html} \citep{Sneden73}, and compared to $\alpha$-element-enhanced stellar atmosphere models ({\sc atlas9}; \citealt{Kurucz93}). For a given {\sc atlas9} model and for each observed line, {\sc moog} returns an abundance ($\epsilon$), and a deviation of that abundance from the model ($\delta\epsilon$).

Models were tuned using a nested, iterative process, to derive three atmospheric parameters: $v_{\rm t}$, $T_{\rm eff}$ and [Fe/H]. First, $v_{\rm t}$ was adjusted to minimise the correlation between equivalent width and $\epsilon$. Secondly, $T_{\rm eff}$ was adjusted to minimise the correlation between excitation potential and $\epsilon$, then $v_{\rm t}$ adjusted accordingly. Finally, the model [Fe/H] is adjusted to match the average $\epsilon$(Fe), and log($g$), $T_{\rm eff}$ and $v_{\rm t}$ are adjusted to remove any trend in their respective parameters. Solar abundances from \citet{AGSS09} were used to convert the absolute ($\epsilon$) abundances to relative (square-bracket notation) abundances.

An attempt was then made to calculate the star's log($g$), by balancing the abundances $\epsilon$(Fe {\sc i}) and $\epsilon$(Fe {\sc ii}). However, only three Fe {\sc ii} lines were measurable in the data, and it was not possible to find a unique, stable solution with positive log($g$). Negative log($g$) is not permitted by the {\sc atlas9} models. For a typical AGB star on the thermally pulsating AGB (TP-AGB), we expect log($g$) $\sim$ 0 dex, and \citet{UGT16} suggest log($g$) $\approx$ 0.18 dex.

Assuming log($g$) = 0 dex, a fit is found for $T_{\rm eff} = 3620$ K, with [Fe/H] = --1.15 dex and $v_{\rm t} = 1.59$ km s$^{-1}$. These values are relatively stable against excursions to higher log($g$): for example, setting log($g$) = 0.18 dex yields $T_{\rm eff} = 3620$ K, with [Fe/H] = --1.08 dex and $v_{\rm t} = 1.58$ km s$^{-1}$. Exact constraint on the metallicity is not possible, but a range of [Fe/H] = --1.3 to --1.0 dex can be estimated, based on the plausible values for the other three parameters. The standard deviation in iron abundance among the 37 Fe {\sc i} lines in the final best fit is 0.278 dex.

\subsubsection{A visual check}
\label{SpecAtmosVisSect}

In addition to this fit, a visual comparison was performed, comparing the observed spectrum to the {\sc bt-settl} model atmosphere spectra \citep{AGL+03}. Fitting of the 647 nm TiO bandhead strongly suggest that $T_{\rm eff} \gtrsim 3700$ K and/or [Fe/H] $\lesssim$ --2 dex (Figure \ref{SpecFig}). This is in contrast to the above analysis and the fitting by \citet{UGT16} of the 705 nm TiO bandhead. This likely reflects problems with dynamical and three-dimensional effects, not included in the models (cf.\ \citealt{LNH+14}).

We conclude that the temperature of 3620 K derived above (and the temperature of $3634 \pm 20$ K derived by \citet{UGT16}) are approximately correct, but that the metallicity is significantly higher than their value, and closer to [Fe/H] $\sim$ --1.15 $\pm$ 0.15 dex. Precise estimation is difficult, given non-LTE effects, and we remind the reader that a single spectrum provides an instantaneous measure of a parameter (e.g., temperature) that may vary considerably throughout a star's pulsation cycle.

A temperature of 3620 K would not give a clear warming compared to literature spectra. Spectral types of M2--M4 have been estimated for this star in spectra taken between 1897 and 1958 \citep{TCC28,LBH+43,Keenan66}. A Morgan--Keenan spectral type of M2--M4 corresponds to temperatures of $T_{\rm eff} \approx 3574-3736$ K \citep{FPT+94}. The features by which these spectral types are measured are not listed explicitly hence, since spectral standards are typically solar-metallicity stars, they should be interpreted with some caution. However, if the spectral type has not been cooler than M4 in recorded history, \citep{FPT+94} implies the stellar temperature cannot have been below $\sim$3574 K. Based on its current temperature of $\sim$3620 K and allowing for measurement errors, it cannot have increased by more than $\sim$100 K since the period of stability before 1955. Assuming little temperature change, the $\sim$18 per cent decrease in radius implied from the period change (Section \ref{PropBackSect}) converts via $L \propto R^2 T^4$ to a decrease in luminosity of $\gtrsim$36 per cent since 1955. This is in line with the expectations of a star entering a thermal pulse (Section \ref{MESASect}), but we remind the reader that this does not account for changes in the stellar interior structure, so should only be taken as indicative of likely changes.

To summarise our spectroscopic findings and our previous discussion on its distance, we identify the following parameters for RU Vul: $d \approx 2000 \pm 165$ pc, $T \approx 3620 \pm \sim 100$ K, $\log(g) \approx 0.0 \pm \sim 0.2$ dex, [Fe/H] = --1.15 $\pm$ $\sim$0.15 dex.

\subsection{Atmospheric parameters from photometry}
\label{PhotAtmosSect}

\subsubsection{SED fitting}
\label{SEDSect}

For comparison, the spectral energy distribution (SED) of RU Vul (Figure \ref{SEDFig}) was fit with the code of \citet{MvLD+09}, with the additions of \citet{MZB12} and \citet{MZW17}. To ensure a fully independent derivation from the spectroscopic determination, we do not use the metallicity, temperature or surface gravity derived in the fit. Instead, we assume the original [Fe/H] = --1.6 dex from \citet{UGT16}. To set the model's surface gravity, we further assume $M$ = 0.6 M$_\odot$ and $d = 2$ kpc. We also assumed $E(B-V) = 0.096$ mag \citep{SF11} and a \citet{Draine03} reddening law. The model is not sensitive to these exact choices, and we examine the sensitivity of the fit to these assumptions below.

A larger uncertainty derives from the stellar variability. The full range of literature data on RU Vul ranges from $U$-band (0.38 $\mu$m) to 60 $\mu$m. Since RU Vul is a variable star, the final temperature is very sensitive to the input photometry, particularly at bluer wavelengths. These data better constrain the stellar temperature but are more subject to variability: for example, using different epochs of $UBV$ photometry from \citet{Koester74} results in temperature changes of up to $\pm$130 K. Redder wavelengths ($\gtrsim$3.4 $\mu$m) are affected by emission from circumstellar dust. Consequently, it is important to fit the SED using time-averaged photometry in the optical, and avoid going too far into the infrared such that infrared excess from dust emission dominates.

Fitting the data with photometry restricted to the \emph{Hipparcos} mean magnitude \citep{vanLeeuwen07} and the 2MASS $JHK_{\rm s}$ magnitudes \citep{SCS+06} returns a temperature of 3639 K, which we expect is accurate to within $\pm\sim$100 K (cf.\ \citealt{MJZ11,CMK16,MZW17}). For the assumed $M = 0.6$ M$_\odot$ and $d = 2$ kpc, this would equate to $L = 3125 \pm 516$ L$_\odot$ with $\log(g) = -0.08 \pm 0.07$ dex. The assumptions we made have little effect on this model: assuming [Fe/H] = --1.0 dex raises the fitted parameters by 27 K and 13 L$_\odot$; assuming $M = 0.9$ M$_\odot$ has no effect on temperature but decreases the luminosity by 7 L$_\odot$; halving the interstellar reddening contribution decreases the parameters by 49 K and 137 L$_\odot$.

Reddening by \emph{circumstellar} dust may mean that both temperature and luminosity are depressed from the true values. However, the close agreement with the spectroscopic temperature (3639 K versus 3620 K) suggests that reddening by circumstellar dust is negligible. \emph{Quantitively} addressing this is not easily possible with these data. \emph{Qualitatively} addressing all the sources of error, we estimate the temperature from SED fitting cannot be much more than 100 K below the spectroscopic temperature. While it is difficult to numerically translate this to an optical depth (as it does not apply to a specific wavelength), a $\sim$100 K offset should be produced by an optical depth of $\tau(1.6\,\mu{\rm m}) \approx 0.02$ mag, approximately equivalent to a $V$-band optical depth of $\tau_V \approx 0.1$ mag (depending on the dust optical properties). Hence, we can estimate that $\tau_V \lesssim 0.1$ mag for RU Vul, despite its unusually large $K_{\rm s}-[22]$ colour. Note that this does not include any grey opacity resulting from larger, micron-sized grains, though we would expect some extinction from their precursor smaller grains if such large grains existed.

To summarise this discussion, we determine the luminosity of RU Vul to be $L = 3125 \pm 516$ L$_\odot$, and confirm the previously calculated $T \approx 3620 \pm {\sim}100$ K and $\log(g) \approx 0.0 \pm {\sim}0.2$ dex. Via $L \propto R^2 T^4$, this therefore provides $R_\ast = 142 \pm 14$ R$_\odot$ (0.66 $\pm$ 0.07 AU).

\subsubsection{Comparison to $\omega$ Cen LEID 42044}
\label{LEID42044Sect}

A comparison can be drawn with stars in the globular cluster $\omega$ Cen ([Fe/H] $\approx$ --1.62, \citealt{Harris10})\footnote{Other clusters are a closer metallicity match, but are not populous enough to host many AGB stars.}. It is clear that the highly evolved stars in $\omega$ Cen are both less dusty and much cooler than RU Vul. The closest match is LEID 42044 (V186 in \citet{CMD+01}; $T_{\rm eff} = 3708$ K; \citealt{MvLS+11}), a semi-regular variable with a visual peak-to-peak amplitude of 0.5 mag (cf.\ RU Vul's amplitude of 0.39 mag). The optical spectrum and normalised SED of LEID 42044 \citep{vLSM+09} closely match those of RU Vul (Figure \ref{SEDFig}). Aside from their differing dust production, these appear very similar stars in their SEDs, spectra and amplitude of variability. This emphasises the portrayal of RU Vul as a star with significant infrared dust emission, but without comparable optical reddening by said dust.

\subsection{Stellar evolution modelling}
\label{MESASect}

\begin{figure*}
\centerline{\includegraphics[height=0.47\textwidth,angle=-90]{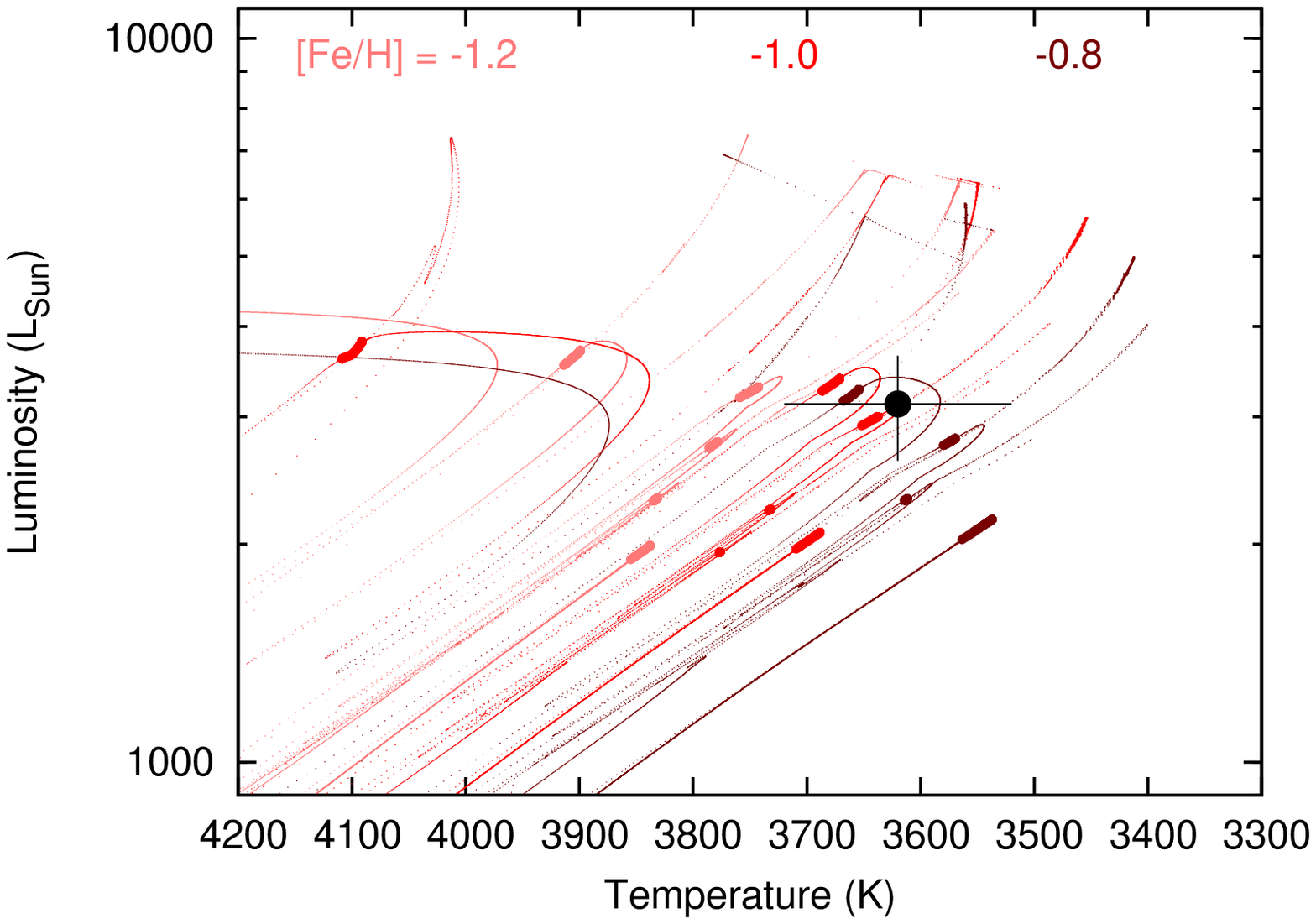}
            \includegraphics[height=0.47\textwidth,angle=-90]{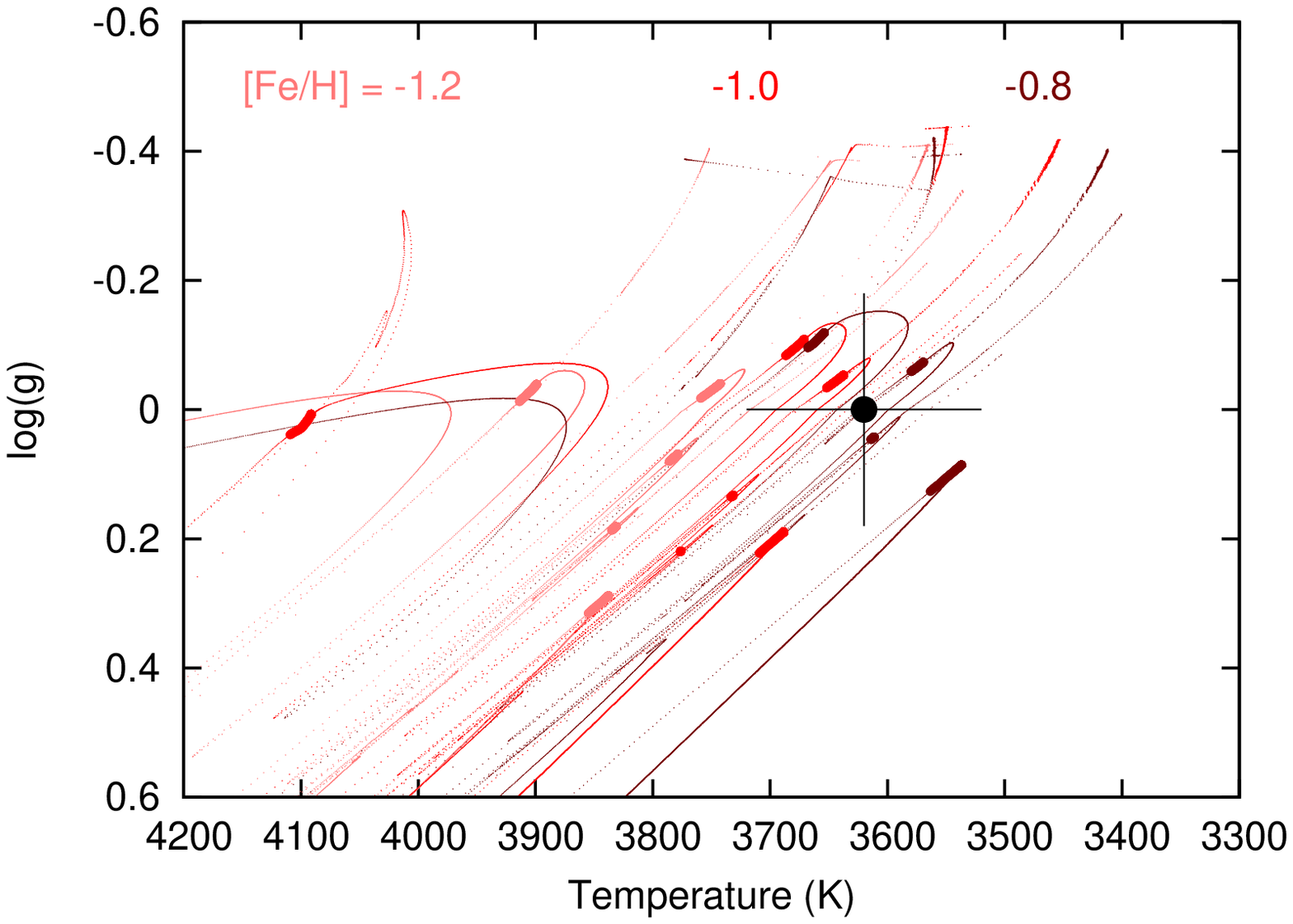}}
\centerline{\includegraphics[height=0.47\textwidth,angle=-90]{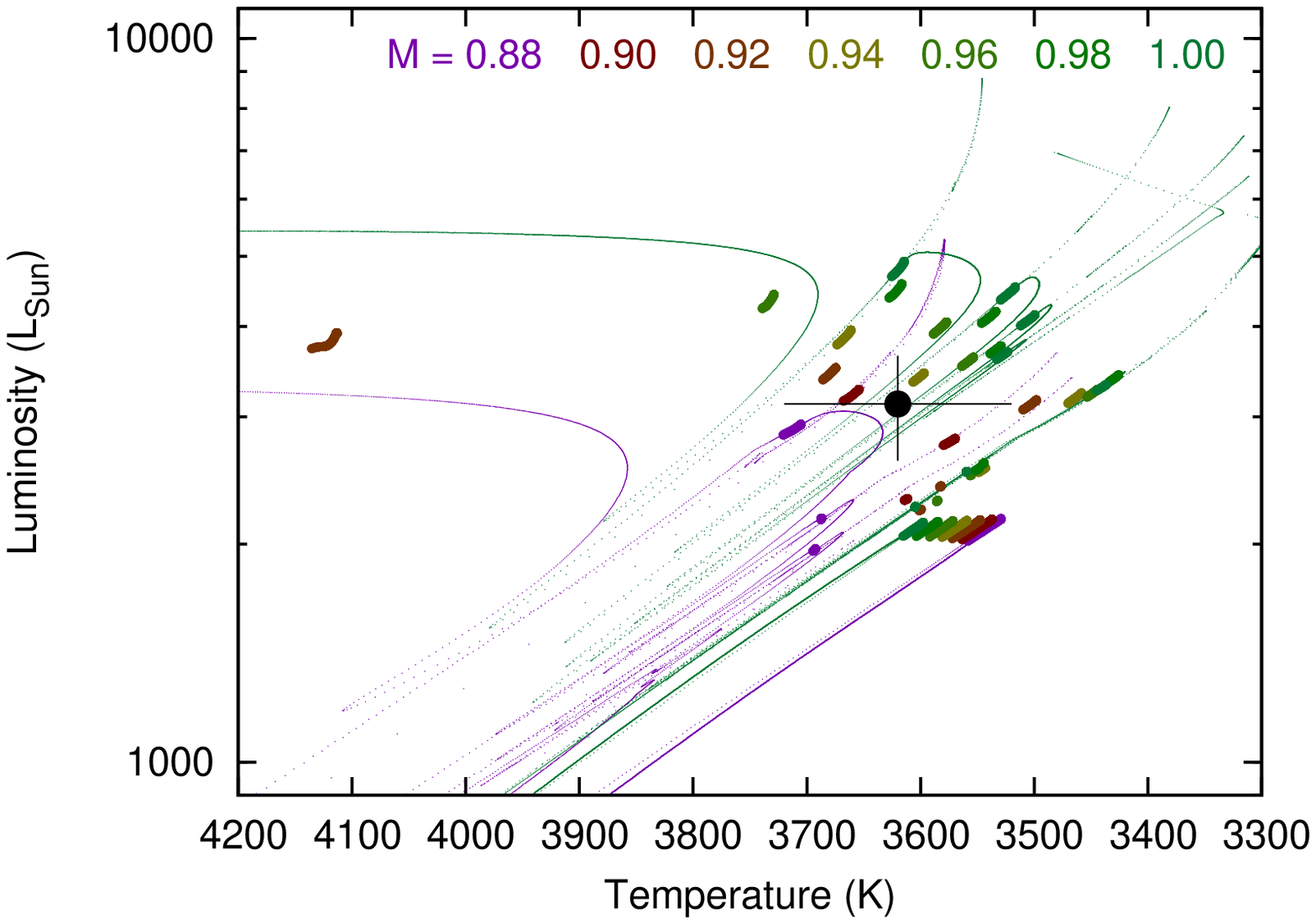}
            \includegraphics[height=0.47\textwidth,angle=-90]{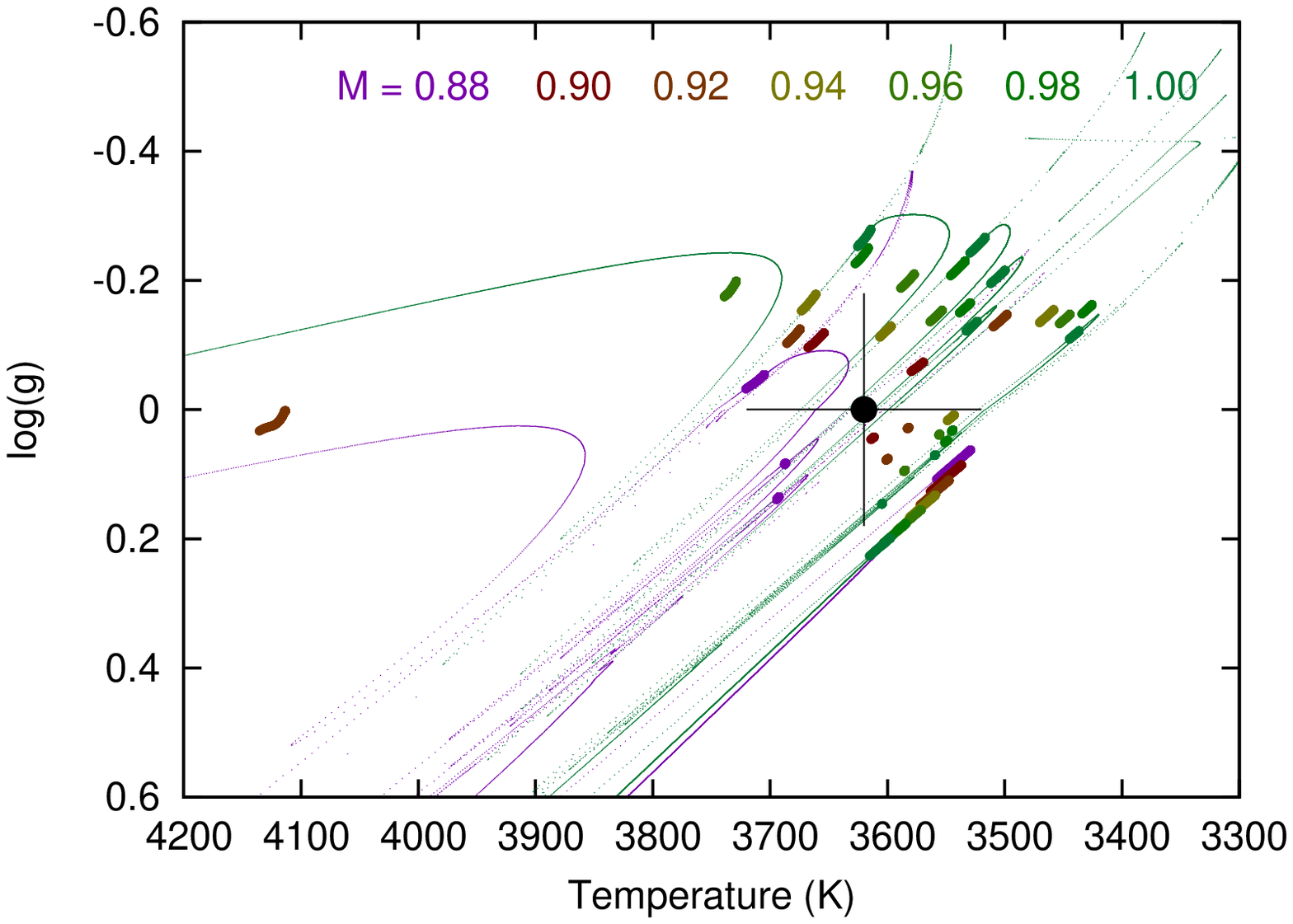}}
\caption{Photometric (left) and spectroscopic (right) Hertzsprung--Russell diagrams, showing the approximate position of RU Vul with approximate error bars (black point). The coloured points show {\sc mesa} stellar evolution tracks from \citet{MZ15b}. Larger points show where models undergo maximum helium-burning luminosity at the start of a thermal pulse (including the helium flash at the RGB tip), approximately representative of RU Vul in 1955. Top panels show how the evolutionary tracks vary with metallicity (left to right / light to dark; [Fe/H] = --1.2, --1.0 and --0.8 dex at 0.90 M$_\odot$). Bottom panels show pre-computed models from \citet{MZ15b} at [Fe/H] = --0.80 dex but with different masses (bottom to top; $M$ = 0.88 to 1.00 M$_\odot$). Only the most- and least-massive tracks are shown in full.}
\label{HRDFig}
\end{figure*}

To ensure our results for RU Vul match with evolutionary theory, we compared our observed properties of RU Vul to {\sc mesa} \citep{PBD+11,PCA+13} stellar evolution models computed for \citet{MZ15b}. These models were designed to accurately represent the late-stage evolution of globular cluster stars, so should be applicable for the chemically and evolutionarily similar stars of the Galactic halo and thick disc. These models have considerable sensitivity to the parameterisation employed for their mass loss, however \citet{MZ15b} calibrated the mass-loss formulism on similar globular cluster stars, so this should not introduce large errors. The models are additionally sensitive to departures from local thermodynamic equilibrium and the adopted efficiency of convection and additional mixing terms, particularly during the thermal pulses: this is likely to affect the exact temperature of models and their variation in the Hertzsprung--Russell diagram during thermal pulses. Such departures may be expected to be $\sim\pm$100 K \citep[e.g.][]{LNH+14}.

The resulting evolutionary tracks are shown in Figure \ref{HRDFig}. Two parameters are explored: metallicity and mass\footnote{The reader should bear in mind that each of these figures represents a slice through mass--metallicity space, thus it is possible to have a combination of metal-poor models with higher mass. However, bearing in mind the chemical evolution of the Galaxy, more-massive, more-metal-poor stars are less common.}. The increase in stellar temperature at lower metallicities, and the increase in the luminosity of thermal pulses at higher stellar masses can both be seen.

Thermal pulses in the [Fe/H] = --1.2 dex models lie to the warmer side of the observed position of RU Vul, hence the {\sc mesa} models indicate the metallicity of RU Vul should be higher than this. The [Fe/H] = --0.8 dex models are still consistent, because the stars become warmer as they shed their envelopes, but a metallicity this high is inconsistent with the spectroscopic measurement ([Fe/H] = --1.15 $\pm$ $\sim$0.15 dex; Section \ref{SpecAtmosSynthSect}). Consequently, stellar evolution modelling provides evolutionary tracks consistent with the higher-metallicity end of our the spectroscopic metallicity derivation. Nevertheless, due to the strongly out-of-equilibrium nature of this star, we retain the spectroscopic estimate as the likely more-accurate answer.

Meanwhile, {\sc mesa} models at 0.88 M$_\odot$ barely achieve thermal pulses, and the luminosity at the start of those pulses is barely consistent with that of RU Vul. Conversely, in models above $\sim$1.00 M$_\odot$, only the first thermal pulse is faint enough to be consistent with the luminosity of RU Vul\footnote{Subsequent thermal pulses are allowed in models up to $\sim$1.02 M$_\odot$ if the star has faded by the predicted 36 per cent in the last few decades (Section \ref{SpecAtmosVisSect}).}. Given the likely range of initial masses predicted from its location in the Galactic Halo (0.78--0.92 M$_\odot$; Section \ref{PropBackDistSect}), and allowing for uncertainties in the efficiency of stellar mass loss, we predict the initial mass of RU Vul was in the range 0.84--0.92 M$_\odot$.

For low-mass stars, considerable mass loss ($\sim$0.2 M$_\odot$) occurs on the RGB \citep[e.g.][]{Rood73,GCB+10,MJZ11,MZ15b}, with more occurring on the early-AGB. The mass of stars in this mass and metallicity range at the start of the TP-AGB is predicted by our {\sc mesa} models to be 0.59--0.64 M$_\odot$. Population II stars like RU Vul leave a remnant of $M \sim 0.53 \pm 0.02$ M$_\odot$ \citep[e.g.][]{Kalirai13}. Hence we can constrain the mass of RU Vul to $M \approx 0.575 \pm 0.065$ M$_\odot$.

Our findings are summarised in Table \ref{ParamsTable}, which presents our final adopted parameters for RU Vul.

\begin{center}
\begin{table}
\caption{Estimated parameters of RU Vul at present and before the thermal pulse. Bracketed values indicate parameters assumed to be unchanged. Italic figures indicate parameters expected to change from evolutionary and pulsation theory, but which are not directly observed.}
\label{ParamsTable}
\begin{tabular}{l@{}c@{}c@{}c@{}c@{}c}
    \hline \hline
    \multicolumn{1}{c}{Parameter} & \multicolumn{1}{c}{Symbol} & \multicolumn{1}{c}{Current} & \multicolumn{1}{c}{Pre-1955} & \multicolumn{1}{c}{Units} & \multicolumn{1}{c}{Section} \\
    \hline
    Distance     & $d$           & \multicolumn{2}{c}{($2000 \pm 165$)}         & pc        & \ref{PropBackSect} \\
    Temperature  & $T_{\rm eff}$ & $3620 \pm {\sim}100$ & $\lesssim${\it 3620}  & K         & \ref{SpecAtmosSect}, \ref{SEDSect} \\
    Luminosity   & $L$           & $3125 \pm {\sim}516$ & $\gtrsim${\it 4250}   & $L_\odot$ & \ref{SEDSect} \\
    Radius       & $R$           & $142 \pm 14$         & $\approx${\it 168}    & $R_\odot$ & \ref{SEDSect} \\
    Surf.\ grav. & $\log(g)$     & $0.0 \pm {\sim}0.2$  & $\sim$--0.1           & dex       & \ref{SpecAtmosSect}, \ref{MESASect} \\
    Metallicity  & [Fe/H]        & \multicolumn{2}{c}{($-1.15 \pm {\sim}0.15$)} & dex       & \ref{SpecAtmosSect}, \ref{MESASect} \\
    Mass         & $M$           & \multicolumn{2}{c}{($0.575 \pm 0.065$)}      & $M_\odot$ & \ref{MESASect} \\
    Period       & $P$           & 108                  & 158.3                 & days      & \ref{IntroSect},\ref{PropBackSect} \\
    \hline
\end{tabular}
\end{table}
\end{center}



\section{The gaseous wind of RU Vul}
\label{ALMASect}

\subsection{ALMA observations}
\label{ALMAObsSect}

\subsubsection{Observational setup}
\label{ALMAObsObsSect}


\begin{figure}
\centerline{\includegraphics[height=0.47\textwidth,angle=-90]{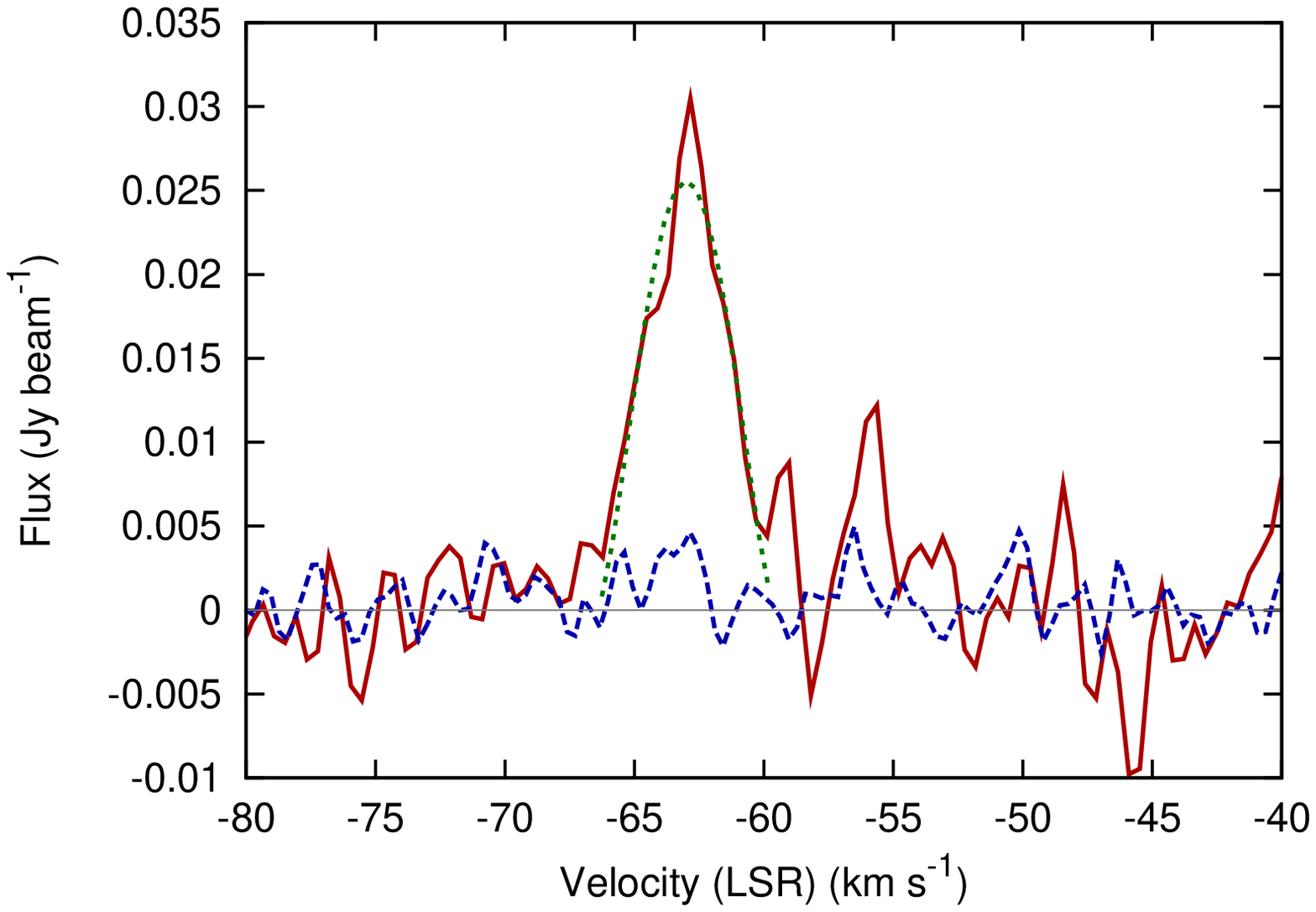}}
\centerline{\includegraphics[height=0.47\textwidth,angle=-90]{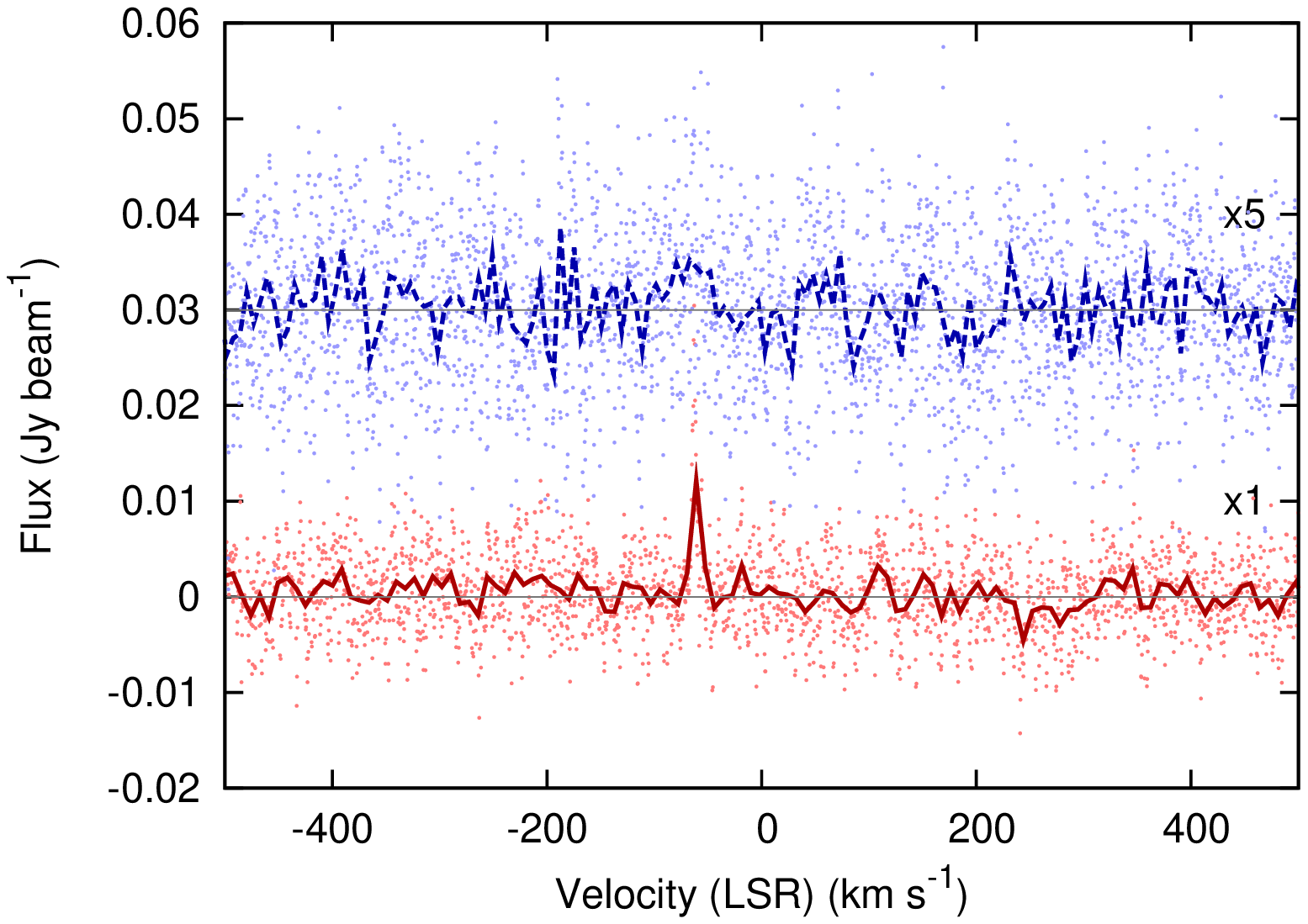}}
\caption{ALMA spectra of RU Vul, in CO (3$\rightarrow$2) (red, solid line) and (2$\rightarrow$1) (blue, dashed line). The thinner, short-dashed, green line shows the modified parabola fit to the CO (3$\rightarrow$2) line. The bottom panel shows a wider section of the spectra, binned by a factor of 20 (to 8.4 and 6.4 km s$^{-1}$, respectively) to show broader but weaker lines (the original points remain). The (2$\rightarrow$1) spectrum has been multiplied by a factor of five and offset, for clarity.}
\label{ALMAFig}
\end{figure}

RU Vulpeculae was observed with ALMA in Bands 6 and 7 (230 and 345 GHz) on 2017 Mar 18 and 27, respectively. The C40--1 configuration was used, providing minimum and maximum baselines of 15 and 155 m. For the Band 6 observations, a spectral window of bandwidth 937 MHz and channel width 244 kHz (0.42 km s${-1}$) was placed on the $^{12}$C$^{16}$O $J$=2$\rightarrow$1 line (230.59 GHz), and three continuum windows with bandwidth 1.7 GHz were placed at 214.11, 215.98 and 229.19 GHz. J1751+0939 was used as a bandpass calibrator, J2051+1743 was used as a phase reference, and Titan was used as a flux calibrator. For the Band 7 observations, a spectral window (1.87 GHz / 488 kHz; 0.32 km s$^{-1}$) was placed on the $^{12}$C$^{16}$O $J$=3$\rightarrow$2 line (345.89 GHz), with three continuum windows (bandwidth 1.7 GHz) placed at 347.84, 333.97 and 335.84 GHz. J2148+0657 was used as a bandpass calibrator, J2039+2152 was used as a phase reference, and Neptune was used as a flux calibrator.

Data were reduced using the automated pipeline (CASA version 4.7) and the resulting circular maps cover a radius of 20$^{\prime\prime}$ and 12.5$^{\prime\prime}$ from the central star, for Bands 6 and 7, respectively. The synthesised beam size is 2.1$^{\prime\prime}$ $\times$ 1.7$^{\prime\prime}$ at $J$=2$\rightarrow$1 and 1.6$^{\prime\prime}$ $\times$ 1.0$^{\prime\prime}$ at $J$=3$\rightarrow$2. Visual inspection of the archived data products indicated no re-reduction of the data was necessary.

The resulting line observations have noise levels ($\sigma$) of 2.2 and 4.1 mJy beam$^{-1}$ channel$^{-1}$ for $J$=2$\rightarrow$1 and 3$\rightarrow$2, as measured in regions near their respective beam centres. Spectra of both lines are shown in Figure \ref{ALMAFig}. Continuum images created from combining all four spectral windows have noise levels of 0.05 and 0.15 mJy beam$^{-1}$, for Bands 6 and 7, respectively.

\subsubsection{Continuum detections}
\label{ALMAObsContSect}

RU Vul is weakly detected as a point source in both the continuum and $J$=3$\rightarrow$2 line emission at the expected position\footnote{cf.\ the \emph{Gaia} DR2 position (projected to epoch 2017.22) 20$^h$38$^m$52$^s$.6868 +23$^\circ$15$^\prime$31$^{\prime\prime}$272} to within measurement errors (the peak CO (3$\rightarrow$2) line flux in Band 7 is found at 20$^h$38$^m$52$^s$.69 +23$^\circ$15$^\prime$31.$^{\prime\prime}$.3). The CO (2$\rightarrow$1) line in Band 6 is only very marginally detected (Section \ref{ALMAObsLineSect}). The continuum flux at 341 GHz is 0.22 $\pm$ 0.05 mJy, and at 222 GHz is 0.085 $\pm$ 0.015 mJy. Estimates of the contributions from different components are listed in Table \ref{CtmTable} and discussed in later sections.

An unrelated point source exists {\rm in the continuum maps,} 3.7$^{\prime\prime}$ to the north (20$^h$38$^m$52$^s$.61 +23$^\circ$15$^\prime$34$^{\prime\prime}$.7), with a peak flux of 0.72 mJy at 341 GHz and 0.31 mJy at 222 GHz, giving a spectral index consistent with a blackbody ($\alpha = 2$, for $F_\nu \propto \nu^\alpha$). This is within the full-width half-maximum (FWHM) of all infrared photometry longward of $K$-band (e.g., the \emph{WISE} FWHM is 6$^{\prime\prime}$--12$^{\prime\prime}$). However, the mid-infrared counterpart is centred on RU Vul itself, even for longest-wavelength mid-infrared data, indicating the mid-infrared emission is associated with RU Vul itself.

\subsubsection{Line detections}
\label{ALMAObsLineSect}

RU Vul is strongly detected in the $J$=3$\rightarrow$2 line, at the expected spatial position, with a line peak at $v_{\rm LSR} = -62.8 \pm \sim0.3$ km s$^{-1}$. This is consistent with the \emph{Gaia} Data Release 2 \citep{GaiaDR2} velocity of $v_{\rm LSR} = -62.38 \pm 0.64$ km s$^{-1}$. The line profile was extracted from the reduced image data cube (Figure \ref{ALMAFig}). The line peaks at a flux of 30.5 $\pm$ 4.1 mJy, and is roughly triangular in shape, with a half-width half-maximum of 1.8 km s$^{-1}$ and a half-width at zero power of $\sim$3 km s$^{-1}$. The integrated intensity over the range --66 to --60 km s$^{-1}$ is $I_{CO(3 \rightarrow 2)}$ = 102 $\pm$ 15 mJy km s$^{-1}$, providing a 6.7$\sigma$ detection.

The integrated intensity in the $J$=2$\rightarrow$1 line over the same --66 to --60 km s$^{-1}$ velocity range is $I_{\rm CO(2 \rightarrow 1)}$ = 10.6 $\pm$ 9.3 mJy km s$^{-1}$, i.e., there is no clear detection. However, the lower panel of Figure \ref{ALMAFig} shows a flux excess at the velocity of the CO (2$\rightarrow$1) line, but spread over a wider velocity range (--63 $\pm$ $\sim$30 km s$^{-1}$). If we assume Gaussian background noise in the rest of the spectrum, the integrated flux over this $\pm$30 km s$^{-1}$ velocity range is 0.106 $\pm$ 0.023 Jy km s$^{-1}$, or a 4.6$\sigma$ detection. This increases to 5.4$\sigma$ for $\pm$20 km s$^{-1}$. If this is a real detection, it would imply a relatively fast, underlying wind at large radii, contrasting with the narrow CO (3$\rightarrow$2) line. No other strong peaks of this characteristic width exist in the spectrum. Despite the statistical prominence of this line, the fact it does not match the width of the $J$=3$\rightarrow$2 line means we are not confident in stating that this line is real and, if it is, what astrophysical origin it might have. We treat it as a probable non-detection in further discussion but allow for its possible contribution to the 222 GHz continuum flux.

The contribution of these lines to the observed continuum flux of RU Vul is listed in Table \ref{CtmTable}.

\begin{center}
\begin{table}
\caption{Summary of ALMA continuum observations.}
\label{CtmTable}
\begin{tabular}{lrrr}
    \hline \hline
    \multicolumn{1}{c}{Contribution} & \multicolumn{2}{c}{Flux ($\mu$Jy)} & \multicolumn{1}{c}{Section} \\
    \  & \multicolumn{1}{c}{341 GHz} & \multicolumn{1}{c}{222 GHz} & \  \\
    \hline
    Total observed       & 220 $\pm$ 50 &    85 $\pm$ 15 & \ref{ALMAObsContSect} \\
    CO lines             &  17 $\pm$  3 &        $<$16 & \ref{ALMAObsLineSect} \\
    Stellar blackbody    &  71 $\pm$  6 &    28 $\pm$  2 & \ref{ALMACtmSect} \\
    Radio photosphere    &  71 $\pm$  6 &    28 $\pm$  2 & \ref{ALMACtmSect} \\
    Remaining dust       &  61 $\pm$ 51 &    29 $^{+15}_{-31}$ & \ref{ALMACtmSect} \\
    \hline
\end{tabular}
\end{table}
\end{center}


\subsection{Wind properties}
\label{WindSect}

Following \citet{DBDdK+10}, we fit a modified parabola of the form:
\begin{equation}
F(v) = F_{\rm max} \left[ 1 - \left( \frac{v - v_0}{v_{\rm exp}} \right)^2 \right]^{\beta/2}
\end{equation}
to the observed $J$=3$\rightarrow$2 spectrum. The best fit provided a peak intensity for the CO line of $F_{\rm max} = 25.6$ mJy, a stellar velocity of $v_0 = -63.02$ km s$^{-1}$, an expansion velocity of $v_{\rm exp} = 3.55$ km s$^{-1}$, with a parabolic fit parameter of $\beta = 3.41$, at a reduced $\chi^2$ of 0.42. This fit is also shown in Figure \ref{ALMAFig}. The formal errors on this fit are not necessarily meaningful, as they do not include errors arising from the assumption of a steady, spherical, homogeneous outflow (we return to this later in Section \ref{DiscSummarySect}).

Converting this intensity to a mass-loss rate is difficult. Different scaling relations (e.g.\ \citealt{RSOL08,DBDdK+10}) provide wildly different mass-loss rates, as they are not calibrated on stars in this regime, nor on observations with telescopes with such small beam sizes as ALMA. Additionally, the implied ratio of $I_{\rm CO(3\rightarrow2)}/I_{\rm CO(2\rightarrow1)} \gtrsim 5$ is much greater than typically found in Galactic stars, even those with optically thin winds (e.g.\ \citealt{DBDdK+10}). We will return to this point in Section \ref{DustMinSect}.

To exemplify these problems, we can create an order-of-magnitude estimate for the mass-loss rate, by scaling stars of known mass-loss rate to the observed parameters of RU Vul. \citet{MZS+16} observed the marginally metal-poor Galactic thick-disc star EU Del. At 2 kpc, EU Del would have $I_{CO(3\rightarrow2)} = 0.50$ Jy km s$^{-1}$ and $I_{CO(2\rightarrow1)} = 0.26$ Jy km s$^{-1}$. However, accounting for RU Vul being $\sim$10$\times$ more metal-poor, hence has an CO:H$_2$ ratio 10$\times$ lower, this would equate to 0.05 and 0.03 Jy km s$^{-1}$, compared to RU Vul's 0.10 and 0.01 Jy km s$^{-1}$. EU Del's mass-loss rate is $\dot{M} = 4.7^{+5.3}_{-3.7} \times 10^{-8}$ M$_\odot$ yr$^{-1}$, and this implies that RU Vul has $\dot{M} \sim 10^{-8}$ to $10^{-7}$ M$_\odot$ yr$^{-1}$, depending on the exact method of determination and its absolute accuracy.


\section{The dust around RU Vul}
\label{DustSect}

\subsection{VISIR observations}
\label{VISIRSect}

An $N$-band spectrum ($R = \delta \lambda / \lambda \approx 350$) of RU Vul was taken with the upgraded VISIR (Very Large Telescope Imager and Spectrometer for the mid-Infrared) spectrograph on the European Southern Observatory's (ESO's) Very Large Telescope (VLT; programme 099.D-0201(A)), on the morning of 2017 June 01. A comparison observation of the K5 giant star HD 194934 then followed. The ESO {\sc reflex} pipeline version 4.3.1\footnote{\url{https://www.eso.org/sci/software/pipelines/visir/visir-pipe-recipes.html}} was used to extracted the spectrum and calibrate it in wavelength. The observation of HD 194934 was used to corrected the spectrum of RU Vul for telluric lines, and to flux calibrate it. The airmass difference between the two observations (1.573--1.844 and 1.950--2.063, respectively) led to significant remnant telluric features in the final spectrum of RU Vul, so we perform additional telluric and flux calibrations below.

\begin{figure}
\centerline{\includegraphics[height=0.45\textwidth,angle=-90]{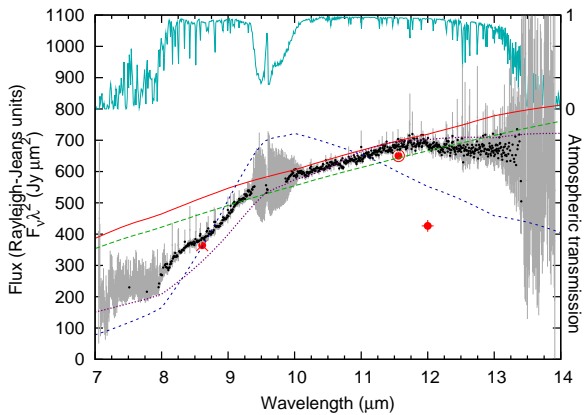}}
\caption{The VISIR spectrum of RU Vul (black points with grey error bars). Photometric points representing the remainder of the SED are shown in red (see also Figure \ref{SEDFig}). The SED has been multiplied by $\lambda^2$ to convert the spectrum into units where the Rayleigh--Jeans tail of the star's SED is horizontal. The VISIR spectrum has been scaled to match the \emph{WISE} 11.3 $\mu$m photometric data point. {\sc Dusty} fits for amorphous carbon (red, solid), metallic iron (green, long dashes), astronomical silicate (blue, short dashes) and mixed-component (magenta, dotted) dust are also shown  (see Section \ref{DustPropSect}). The cyan line at the panel's top shows the terrestrial transmission spectrum. Black data points have been removed from regions of the spectrum that remain heavily affected by atmospheric features.}
\label{VISIRFig}
\end{figure}

The integrated water column during the observations was $\approx$2 mm. The slow chopping frequency of 0.013 Hz meant that the observations ($F_{\rm o}$) are contaminated by remaining telluric lines. To correct for this, a telluric transmission spectrum\footnote{\url{http://www.gemini.edu/sciops/ObsProcess/obsConstraints/atm-models/cptrans_zm_23_15.dat}} ($T$) was divided out of the spectrum via:
\begin{equation}
F_{\rm final} = F_{\rm o}/(1 + T / s)
\end{equation}
where $s$ is a scaling factor. Reasonable fits were found for $s = 4 \pm 1$, and $s = 4$ was adopted in the final fit. The final spectrum of RU Vul ($F_{\rm final}$) is shown in Figure \ref{VISIRFig}, scaled to the flux density of the 11.3-$\mu$m \emph{WISE} data point.

The VISIR spectrum still shows remnant noise from narrow terrestrial water features ($<$8 and $>$12 $\mu$m) and ozone absorption (9.5 $\mu$m), marked in grey in Figure \ref{VISIRFig}. Between $\sim$10 and 12 $\mu$m, the spectrum is remarkably   straight, but falls off to both the long- and short-wavelength sides, suggesting an unusually broad silicate feature. A weak inflection around 9 $\mu$m could be due to a combination of absorption by SiO at 8 $\mu$m and increasing emission due to the Si-O-Si vibrational band towards $\sim$9.7 $\mu$m \citep[cf., e.g.][]{WOK+11}. The cause of the inflection around 11.9--12 $\mu$m is unknown. Consequently, there is relatively little we can say about the mineralogy of the wind except that it appears dominated by a continuum contribution, with a weak and very broad feature on top that could be attributable to silicate dust.

\begin{figure}
\centerline{\includegraphics[height=0.45\textwidth,angle=-90]{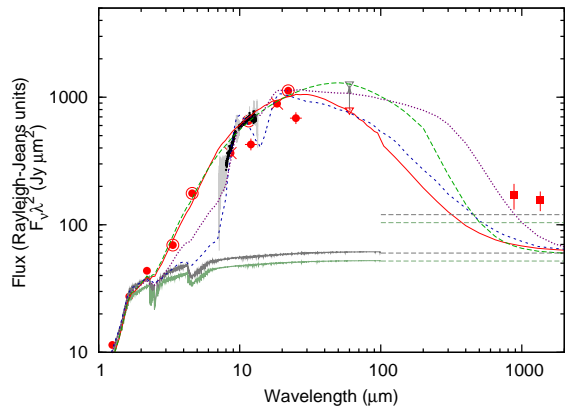}}
\caption{As Figure \ref{VISIRFig}, showing the entire near- and mid-infrared portion of the SED.  The same indicative stellar photospheres at 3600 and 3700 K are given as in Figure \ref{SEDFig}, projected horizontally by the flux units used. The additional horizontal lines a factor of two above these represent the expected radio photosphere (Section \ref{ALMACtmSect}).}
\label{VISIRSEDFig}
\end{figure}


\subsection{ALMA continuum observations}
\label{ALMACtmSect}

The continuum detections in Section \ref{ALMAObsContSect} reflect the contribution of several components. To obtain the emission from cold dust around RU Vul, we must first subtract the contributions from line emission, the stellar blackbody (optical photosphere) and chromospheric emission (radio photosphere). These components are separated here, and in the other sections mentioned in Table \ref{CtmTable}.

Fitting the stellar photosphere with a model atmosphere suggests the star's Rayleigh--Jeans tail is maintained at $\sim$55 $\pm$ 5 Jy $\mu$m$^2$ (Figures \ref{SEDFig} and \ref{VISIRSEDFig}). At 341 and 222 GHz (880 and 1350 $\mu$m), this translates to an expected flux density for a naked photosphere of $\sim$71 $\pm$ 6 $\mu$Jy and $\sim$30 $\pm$ 3 mJy. Even after subtracting the (small) contribution from line emission, the ALMA continuum flux densities are a factor of $\sim$2--3 above these expectations (Table \ref{CtmTable}).

A factor of $\sim$2 is fairly typical for the radio photospheres of mass-losing stars, thought to be caused by H$^-$ and H$_2^-$ free-free interactions in the stellar envelope \citep{RM97}. Observations of these stars at millimetric wavelengths show they indeed appear to have physically larger radii in the millimetre and radio than the optical \citep{OGKH+17}. We indicate the expected factor-of-two increase in flux from the radio photosphere by the second set of horizontal lines in Figure \ref{VISIRSEDFig}.

Approximating the flux from the radio photosphere as identical to the optical photosphere, and subtracting both quantities from the line-corrected continuum flux, we arrive at the contribution to the ALMA continuum flux of cold dust: $61 \pm 51$ $\mu$Jy at 341 GHz and 29 $^{+15}_{-31}$ $\mu$Jy at 222 GHz. For comparison, the flux from the low-mass-loss-rate star R Dor ($K_{\rm s}-[22] \sim 1.3$ mag; $d$ = 59 pc; $F_{\rm dust,\ 341\ GHz} = 67$ mJy; \citep{DRD+18}) would be $\sim$58 $\mu$Jy at the distance of RU Vul \citep{DRD+18}. Consequently, despite having very little flux emitted by cold dust, this is consistent with other low-mass-loss-rate stars.


\subsection{Dust properties}
\label{DustPropSect}

\subsubsection{Fitting the SED}
\label{DustSEDSect}

In Section \ref{PropBackDiscSect}, we proposed that RU Vul is undergoing rapid dust formation. To explore the properties of RU Vul's dusty wind, we perform radiative-transfer modelling using the {\sc dusty} code \citep{INE99}\footnote{\url{http://faculty.washington.edu/ivezic/dusty_web/}}. In unresolved data, the setup for this code relies on a number of assumptions about the dust, its properties, and its distribution. The resulting fits are heavily parameterised, and parameters are highly correlated, meaning a fit is best achieved ``by eye''. It also assumes that dust is formed in an instantaneous thin shell, that dust mineralogy and grain properties remain constant throughout the stellar envelope, and that temperature equilibrium remains between the dust and surrounding gas. Much of this may not be realistic (e.g., \citealt{BH12,BHAE15}). Consequently, the results should only be used indicatively.

The same initial setup was used as for the aforementioned globular cluster stars \citep{MvLD+09,MBvL+11,MvLS+11}. The wind was modelled using a purely radiation-driven wind ({\sc dusty} option {\tt density type = 3}). The 3600 K, log($g$) = 0 dex, [Fe/H] = --1.0 dex {\sc bt-settl} model atmosphere \citep{AGL+03} was used as an input, spectrally degraded to approximately twice the resolution of underlying optical constants (real and imaginary parts to the dust refractive index). Three different sets of optical constants were used to model the wind: amorphous carbon \citep{Hanner88}, metallic iron \citep{OBA+88} and `astronomical' silicates \citep{DL84}. Amorphous carbon and metallic iron are chosen to produce the continuum flux in excess of the stellar photosphere; silicates are chosen to reproduce the 10-$\mu$m emission. The physicality of these choices is discussed in Section \ref{DustMinSect}. We use the standard \citet{MRN77} size limits ($a$ = 0.005--0.25 $\mu$m) and slope ($N(a) \propto a^{-3.5}$).

The dust is modelled as a thick shell with inner and outer radii of $R_{\rm in}$ and $R_{\rm out}$. Parameters specific to each dust component include the dust temperature at the inner edge of the dust envelope ($T_{\rm dust,in}$), which has the effect of moderating the wavelength of the short-wavelength side of the dust distribution and setting the dust inner radius; the ratio $R_{\rm out}/R_{\rm in}$, which (in tandem with $T_{\rm dust,in}$) moderates the long-wavelength side of the distribution; and the optical depth of the models at 0.55 $\mu$m ($\tau_V$), which sets the amplitude of the infrared dust excess compared to the underlying stellar photosphere. The remaining factors are set as follows to provide the fits shown in Figures \ref{VISIRFig} \& \ref{VISIRSEDFig}:
\begin{itemize}
\item {\it Amorphous carbon:} $T_{\rm dust,in} = 750$ K, $R_{\rm out}/R_{\rm in} = 100$ and $\tau_V = 0.65$ mag.
\item {\it Metallic iron:} $T_{\rm dust,in} = 675$ K, $R_{\rm out}/R_{\rm in} = 33$ and $\tau_V = 1.1$ mag.
\item {\it Astronomical silicates:} $T_{\rm dust,in} = 750$ K, $R_{\rm out}/R_{\rm in} = 1000$ and $\tau_V = 1$ mag.
\end{itemize}
Conversions of these factors to physical values are discussed in Section \ref{DustDriveSect}.

A fourth fit was produced with a mixture of dust types, mainly silicates and metallic iron, but also including porous Al$_2$O$_3$. Constants for Al$_2$O$_3$ were taken from \citet{BDH+97}. These approximately reproduce the 12-$\mu$m inflection seen in the spectrum. Components of this wind were fit as: 80 per cent metallic iron, 13 per cent silicate and 7 per cent Al$_2$O$_3$. Other properties are: $T_{\rm dust,in} = 600$ K, $R_{\rm out}/R_{\rm in} = 50$ and $\tau_V = 0.37$ mag. These represent the emission properties of the wind under the following assumptions: that gas and dust are thermally coupled, that grains of each type come from the same \citet{MRN77} size distribution of grains. The model still under-predicts the flux between 2 and 10 $\mu$m but performs better when reproducing the longer-wavelength data.

A unique property of these fits is that $T_{\rm dust,in}$ is considerably cooler than the condensation temperature of these dust species (fits to the dust of other, similiarly metal-poor stars provide values closer to the expected $\sim$1000 K; cf., \citealt{BMvL+09,MvLS+11}). This is consistent with the apparent cooling of dust mentioned in Section \ref{PropBackDiscSect}, and we suggest interpretations of these events in Section \ref{DiscWindSect}.

\subsubsection{Dust mineralogy}
\label{DustMinSect}

{\it Amorphous carbon} creates a featureless $N$-band continuum. However, RU Vul is an oxygen-rich star. \citet{HA07} suggest amorphous carbon can form if UV light dissociates CO, creating free carbon. While RU Vul is far from major sources of UV radiation and does not exhibit an obvious UV excess, the central star is warm enough to emit significant UV radiation. The CO (3$\rightarrow$2) detection shows a substantial amount of CO must remain intact, though the high $I_{\rm CO(3\rightarrow2)}/I_{\rm CO(2\rightarrow1)}$ ratio could indicate CO destruction or abnormal warming at large radii. On balance, we consider it chemically difficult to form amorphous carbon dust around RU Vul.

{\it Metallic iron} has been posited as a common dust species by several authors, both in nearby stars and other evolved objects \citep{KdKW+02,VvdZH+09,MZS+16}, and around globular cluster AGB stars \citet{MSZ+10,MBvLZ11,MvLS+11}. RU Vul is qualitatively similar to the globular cluster stars (Section \ref{LEID42044Sect}), the primary differences being they receive much more UV radiation (e.g. \citealt{MZ15a,MBG+19}) and are less dusty than RU Vul. Conceptually, iron is missing from the gas phase in many locations, including AGB-star winds \citep[e.g.][]{MH10}. However, its high opacity means it should only condense at large radii ($\sim$10 R$_\ast$). If it can do so, and is in radiative equilibrium with the star (rather than thermal equilibrium with the surrounding gas as {\sc dusty} assumes), then it will retain a temperature\footnote{Note that in Section \ref{DustSEDSect} we obtain an inner temperature of $\sim$675 K. The difference from the 1000 K quoted here could be indicative of several scenarios, including that the dust has not yet reached equilibrium with its surroundings due to the changing star, and/or that the wind is not well-represented by pure metallic iron dust.} of $\sim$1000 K. The arguments for this were put forward by \citet{BH12,BHN+13} who modelled that iron-poor silicates can form near the star, while iron-rich silicates can form further out.

The chemical formation of metallic iron in this process is unclear. Iron-rich silicates would normally condense before metallic iron \citep[e.g.][]{GS99,BH12}. \citet{MDADC+19} propose silicate production in hot-bottom-burning stars can be inhibited if oxygen and magnesium abundances are lowered. Metallic iron then forms. While the low mass of RU Vul means hot-bottom-burning cannot be active, this process relies only on creating an environment where the Fe/O ratio is $\gtrsim$1. \citet[][their table 3 and section 3.5]{MBvLZ11} discuss the varying atomic abundances in the wind of an $\alpha$-element-enhanced star. The Fe/O ratio following the condensation of CO and silicate dust is expected to be $\approx$1/4 if the silicate condensate is the enstatite (MgSiO$_3$) end member (and potentially higher if the condensate is closer towards the forsterite [Mg$_2$SiO$_4$] end member). This does not account for any oxygen deposition in alumina dust or water, or where CO$_2$ condenses instead of CO. Since all three species are seen in the infrared spectra of metal-poor, oxygen-rich stars \citep[e.g.][]{MvLS+11}, it should not be difficult to achieve Fe/O $>$ 1 by the $\sim$10 R$_\ast$ formation radius of metallic iron. Consequently, metallic iron could form preferentially to astronomical silicates.

Iron at large radii (e.g., $\sim$10--15 R$_\ast$) may still be at the same temperature as silicates at smaller radii (e.g., $\sim$2--3 R$_\ast$), thus the hot component of metallic iron dust could form $\sim$5 times as much of the wind as the silicates. This would allow metallic iron to dominate the opacity of the wind, even if it represents a comparatively small fraction of the dust. We therefore favour metallic iron as the underlying continuum dust opacity source for RU Vul over amorphous carbon.

{\it Astronomical silicates} are the expected condensate around oxygen-rich AGB stars \citep[e.g][]{GS99}. The \citet{DL84} optical constants used here are an empirical reflection of the dust from Galactic sources, hence may include species contributing to the continuum, like those above \citep[e.g.][]{KdKW+02}. Despite this, a silicate wind cannot alone reproduce the strong dust emission in the 3--8 $\mu$m range, nor the comparative flatness of the $N$-band spectrum longwards of 10 $\mu$m. However, the apparent presence of a 10-$\mu$m silicate feature suggests some silicates are present. Scattering by large silicate grains have been proposed as a wind-driving mechanism (Section \ref{IntroMdotSect}; \citealt{Hoefner08}). While the 10-$\mu$m emission feature may be damped if \emph{only} large silicate grains exist, it is conceptually difficult to grow these without a much more numerous population of much smaller grains to form them from (which would themselves create a 10-$\mu$m emission feature; \citealt{MSZ+10}). Scattering by grains is also expected to create line-profile asymmetries in the optical spectrum \citep{RL81}. These are not seen in RU Vul.

Possible explanations for the excess infrared flux are a thick molecular layer, but it hard for this (likely water-based) layer to avoid showing strong features in the infrared spectrum; free-free emission from circumstellar plasma can also produce continuum emission with a rising spectral index, but requires plasma column densities that cannot be realistically created around AGB stars \citep{MSZ+10}. With lack of a viable alternative, we proceed with the hypothesis that the material responsible for the infrared excess of RU Vul is largely metallic iron with some contribution from astronomical silicates and possibly aluminium oxide.

\subsubsection{Radial dust geometry and wind driving}
\label{DustDriveSect}

\begin{center}
\begin{table*}
\caption{Indicative expectations for temperatures and radii of dust around RU Vul.}
\label{DustTable}
\begin{tabular}{lccccccc}
    \hline \hline
    \multicolumn{1}{c}{Mineralogy} & \multicolumn{1}{c}{$T_{\rm cond}$} & \multicolumn{1}{c}{$R_{\rm cond}$} & \multicolumn{1}{c}{$T_{\rm dust,in}$} & \multicolumn{1}{c}{$R_{\rm dust,in}$} & \multicolumn{1}{c}{$R_{\rm out}/R_{\rm in}$} \\
    \multicolumn{1}{c}{\ } & \multicolumn{1}{c}{(K)} & \multicolumn{1}{c}{(R$_\ast$)} & \multicolumn{1}{c}{(K)} & \multicolumn{1}{c}{(R$_\ast$)} & \ & \multicolumn{1}{c}{(K)} & \multicolumn{1}{c}{(R$_\ast$)} \\
     \hline
    Amorph. C      & 1680 &  4 & 750 & 20 &  100 &  75 &  2000 \\
    Metallic iron  & 1050 & 25 & 675 & 60 &   33 & 118 &  2000 \\
    Ast. silicates & 1100 &  4 & 750 & 20 & 1000 &  24 & 20000 \\
    Mixed          & 1100 &  4 & 600 & 13 &   50 &  85 &   650 \\
    \hline
\end{tabular}
\end{table*}
\end{center}

Since these are the first continuum measurements of a metal-poor star in the far-infrared, they provide the first constraint on the long-wavelength end of the SED. Relatively little emission from cold dust at large radii suggests the wind of RU Vul could be truncated relatively close to the star. While the unresolved nature of the stellar wind in VISIR and ALMA limits the extent of cold dust around RU Vul, VISIR only traces warm ($\gtrsim$300 K) dust and the resolution of ALMA only provides a limit of $R_{\rm outer} \lesssim 0.5^{\prime\prime}$ (1000 AU).

We can use Equation 10 and Figure 1 of \citet{BH12} to derive expected condensation temperatures and radii ($T_{\rm cond}$, $R_{\rm cond}$) for our different dust species. Assuming radiative equilibrium ($T^4 \propto R^2$), we can combine these with our values of $T_{\rm dust,in}$ from Section \ref{DustSEDSect} to obtain $R_{\rm in}$, then use $R_{\rm out}/R_{\rm in}$ to compute the expected extent of the dust shell. These derived quantities are listed in Table \ref{DustTable}. As a reminder, these are indicative values only, particularly for our preferred ``mixed'' dust mineralogy, as the condensation fraction and composition of the dust will change radially in the wind.

While stellar pulsations have the kinetic energy to launch a wind on their own, the shock velocities they impart to the wind are considerably below the escape velocity, so material can be ejected by pulsations to only a few stellar radii \citep[e.g.][]{LWH+05,MvL07,HO18}. Hence, while dust driving can remain effective out to $\sim$100 R$_\ast$ \citep{DJDB+10}, some acceleration must occur within a few R$_\ast$ of the surface. Assuming amorphous carbon is chemically impossible to condense efficiently, a mixed-mineralogy dust wind is the only one that can provide the necessary continuum opacity and condense close to the star.

The idea of a slowly increasing condensation fraction (Section \ref{DustSEDSect}) and slow wind velocity (Section \ref{ALMAObsLineSect}) was modelled in \citet{MBG+19} as being consistent with a \citet{WlBJ+00} ``type B'' wind. Here, the outflow is maintained in a marginal state by a balance between dust condensation (which depends on wind density) and the increased radiation pressure that dust condensation causes (which decreases wind density, hence also decreases dust condensation again). It may be that this state is maintained in RU Vul as well, but the thermal pulse and inhomogeneities in the dust prevent us from stating this conclusively, as we discuss in the next section.

\subsubsection{Optical dust absorption and dust inhomogeneity}
\label{DustGeomSect}

Complicating this picture is the lack of optical absorption that is taking place. Several per cent of the star's light is being reprocessed into the infrared ($\sim$8 per cent is quoted in \citet{MZB12}). This reprocessing should dim the star: our {\sc dusty} fits indicate a $V$-band absorption of $\sim$0.37 mag for our adopted mixed-dust wind, while SED fitting (Section \ref{SEDSect}) indicates $\tau_V < 0.1$ mag.

Reprocessing will also considerably redden the stellar light. This would be reflected in the temperature inferred from fitting the photometric SED, compared to that found by fitting the optical spectrum: these temperatures agree to within uncertainties ($\lesssim$100 K; Section \ref{SpecAtmosSect} \& \ref{PhotAtmosSect}). The expected amount of reddening depends strongly on the type of dust and optical constants used. Were the dust around RU Vul amorphous carbon, the reddening would change the temperature inferred from fitting the photometric SED by up to 1800 K. For our adopted mixed dust model, the offset is much more modest, only $\sim$200 K, but still enough to create a detectable departure in the optical SED. This implies an aspherical or clumpy geometry to the wind that affords us a clear line of sight towards the stellar surface.


\section{The nature of RU Vul}
\label{DiscSect}

\subsection{A summary of the evidence}
\label{DiscSummarySect}

RU Vul is clearly a very peculiar AGB star (we remind the reader of the star's fundamental properties, in Table \ref{ParamsTable}). Its changing period is the reason it was originally targetted by \citet{UvSV+11}, which led to the evidence that it is undergoing the initial dimming phase of a thermal pulse \citep{UGT16}. Its unusually emissive dust envelope is the reason we targetted it, as best metal-poor star detectable by ALMA. We do not have evidence to state whether the thermal pulse and very dusty envelope are linked, but we should expect that the nature of this star's circumstellar material does not reflect either AGB stars or metal-poor stars as a whole. We may therefore expect that a relatively rare phenomenon may be taking place, either temporally rare over the star's entire AGB phase, or rare in terms of star-to-star variation.

The star was stable in observations between the 1920s and early 1960s (Section \ref{PropBackSect}), when the photometric minimum began to get brighter in the visible. Starting in the 1950s, its period began to shorten: indications that the star is shrinking. We may also expect it to get warmer, but the implied warming since 1897 appears limited to $\lesssim 100$ K (Section \ref{SpecAtmosVisSect}). Mid-IR emission from circumstellar dust has increased by around twofold since records began in the 1980s, potentially indicating the onset of a dust-condensation event around the 1950s that is still ongoing (Section \ref{PropBackDiscSect}).

The CO $J$=3--2 emission line has a velocity of only $\sim$1.8--3.5 km s$^{-1}$ (Section \ref{ALMASect}), which includes any turbulent velocity. The roughly triangular shape of the CO (3$\rightarrow$2) velocity profile may be approximated as a Gaussian distribution of internal motions, rather than a net outflow. Therefore, while RU Vul is clearly losing mass in the long term, we cannot \emph{conclusively} state whether the material currently around the star represents an outflowing wind or material orbiting in the circumstellar environment. In the latter case, the possible higher-velocity component is present in CO (2$\rightarrow$1) still allows the possibility of a faster outflow of colder material (e.g., a true wind at larger radii), but its lack of contrast means we cannot draw any conclusions.

The unusually high ratio of $I_{CO(3\rightarrow2)}/I_{CO(2\rightarrow1)}$ (Section \ref{ALMAObsSect}), suggests that either the historical (pre-thermal-pulse) mass-loss rate may have been lower than at present or that freshly ejected material remains bound close to the star. Equally, there is a deficit of very hot dust (Section \ref{DustSEDSect}). Circumstellar dust does not appear to redden the star along our line of sight, suggesting an aspherical dust geometry (Section \ref{DustGeomSect}), limited to radii $\lesssim$1000 R$_\ast$ from the star (Section \ref{DustDriveSect}).

\subsection{Is RU Vul a binary?}
\label{DiscBinarySect}

So far, we have treated RU Vul as an isolated single star. An alternative hypothesis that could explain some of its characteristics is if it is part of an interacting binary system. Roughly half of stars are in binary systems and binary orbital angular momentum can be used to focus material into a disc, if the stars and/or their outflows interact. It is thought that even objects as small as planets may play a role in this process, ultimately shaping asymmetric planetary nebulae \citep[e.g.][]{DeMarco09}. The binary fraction among metal-poor stars appears to be higher than solar-metallicity stars, with up to half of objects hosting a companion \citep{MKB19}, though the exact numbers depend sensitively on the correction for observational biases, and some of these systems will already have formed common envelope systems and merged by the time the star reaches the AGB. Meanwhile, the fraction of stars with giant planets likely decreases to levels of $\sim$a per cent \citep{Adibekyan19}.

Binarity among AGB stars is hard to detect directly. Secondary components cannot normally be detected via their spectroscopic lines due to their lack of contrast against the bright AGB star, and detecting the reflex motion of the AGB star is confounded by the radial pulsations of its atmosphere. Instead, binary companions most obviously manifest themselves in the shaping of the AGB envelope into discs, arcs or spirals, depending on the separation and mass ratio of the system \citep[e.g.][]{KT12,MP12}.

Signatures of binary-induced spirals have been inferred in many of the AGB stars that have been observed at high resolution with ALMA (e.g., \citealt{MMV+12,RMV+14,RMV+17,DRN+15,LZK+18}). These frequencies are consistent with both stellar and planetary-mass companions playing a role in driving the shaping of AGB winds.

Alongside these features, AGB and post-AGB stars often exhibit circumstellar discs. Though rarer, these can form when a binary companion deflects mass loss from the AGB star into an orbital motion, trapping it in the system. This lets significant quantities of gas and dust build up. Like RU Vul, systems with discs tend to be heavily reddened compared to the trendline in the period--infrared-excess diagram (see also Section \ref{DiscPChangeSect}), with colours several magnitudes redder than the nominal $K_{\rm s}-[22] \approx 2$ mag seen for stars in their period rang \citep[cf.][]{MZ16}. Heavily reddened systems can include both edge-on discs like L$_2$ Pup \citep{LKP+15,KHR+16,HRD+17}, and face-on discs like EP Aqr (\citealt{HRD+18}, but see also \citealt{HNTA+19}), hence the extra reddening need not be linked to the obscuration of the star by the circumstellar environment, but the presence of the disc itself. EP Aqr is particularly notable for the broad and narrow velocity components to its CO lines. These stars have similarly large infrared excesses and similar periods to RU Vul. Post-AGB systems with strong infrared excess include IRAS 08544--4431, a binary post-AGB system with an inclined disc with a narrow, triangular CO line profile similar to that of RU Vul \citep{MVWLE+03,Trung09}, and IW Car, a post-AGB star with a rotating disc \citep{BCCA+17}. Given the frequency of interacting companions, and the similarities between the infrared colour and line profiles of interacting systems and those of RU Vul, we must entertain the possibility that the wind of RU Vul is being shaped by an unseen companion.

A relatively face-on disc around RU Vul could present an EP-Aqr- or IRAS 08544--4431-like system, where a narrow component is related to a disk or similar density enhancement in the plane of the sky, while a broader component representing the underlying wind remains undetected (cf., our unclear CO $J=2 \rightarrow 1$ line). The Keplerian velocity of such a disc, as projected into the line of sight, depends on the systemic mass and orbital radius. For a systemic mass of $\gtrsim$0.53 M$_\odot$ and inner radius of $\sim$13 R$_\ast$ (8.5 AU), the orbital speed is $\gtrsim$7.5 km s$^{-1}$. If the CO $J=3 \rightarrow 2$ line comes entirely from a thin disc (i.e., if the density of the disc is sufficiently high compared to the underlying wind), then its full-width, zero-intensity speed of $\sim$3 km s$^{-1}$ requires an inclination angle of $\lesssim$22$^\circ$ from the plane of the sky, which should occur for about one-in-four binary orientations.

A binary companion can be expected to produce radial velocity or astrometric offsets to the AGB star. Radial velocity variations in the plane of the sky should be small for a face-on system. While published literature shows heliocentric radial velocities that vary by $^{+10}_{-14}$ km s$^{-1}$ around the \emph{Gaia} DR2 measurement \citep{DFM95,UvSV+11}, the pulsations of the AGB star, their typical calibration uncertainties ($\sim$5 km s$^{-1}$) and the differences in their reduction method mean we cannot be confident that these represent real motions of the AGB star's centre of mass. Similarly, \citet{KAMT19} identifies a 2.8$\sigma$ anomaly between the \emph{Gaia} position and proper motions projected back to the \emph{Hipparcos} epoch, and the recorded \emph{Hipparcos} position, which could indicate an astrometric shift of the AGB star. However, variations on the stellar surface, or changes in circumstellar scattering or absorption, could cause variations in the \emph{Gaia} proper motions that would replicate this effect. Consequently, while there is a reasonable possibility that a binary companion could be shaping a disc around RU Vul, it is not convincing at this time.

\subsection{The rapidly condensing dust}
\label{DiscWindSect}

The rapid increase in infrared flux (Section \ref{PropBackDiscSect}) is an integral part of understanding the changes to the evolution of the star's wind. Normally, such a change in infrared flux might be tied to an abrupt change in the mass-loss rate \citep[e.g.][]{OIO+15}, and some previous dust-forming events have been linked to ongoing thermal pulses, e.g.: Sakurai's object (V4334 Sgr; \citealt{HJ14}, and references therein), and WISE J180956.27-330500.2 \citep{GYT12}. Material ejected from the optical photosphere around 1955 will, at a wind velocity of 3 km s$^{-1}$, have travelled $\sim$39 AU ($\sim$58 R$_\ast$) from the star, greater than or equal to the modelled inner radius of the dust envelope (13--60 R$_\ast$; Table \ref{DustTable}). An epoch of stronger mass ejection, starting around 1955, could therefore cause the increase in infrared excess that we see. Such an ejection event could ultimately result in a detached shell associated with the thermal pulse.

For RU Vul, there is no clear precursor to trigger this strong increase in the mass-loss rate: the levitating strength of pulsations has declined during the last century, and the bolometric luminosity has likely declined, making it conceptually harder for mass loss to occur. Mass could be suddenly ejected by dumping energy into the stellar atmosphere (e.g., an acoustic pulse associated with the runaway thermonuclear fusion of helium, or by dropping an orbiting planet into the atmosphere), but we might expect this to also cause sudden brightening and disruption to the (stochastically excited) pulsation mechanism over about a free-fall timescale (i.e., less than a pulsation period). Consequently, the idea that RU Vul is experiencing a sudden mass-loss episode is not impossible, but currently has no observational support. Instead, we explore a scenario where the mass-loss rate is constant and changes occur in material that has already been levitated from the stellar surface, but which has not yet condensed into dust.

Evolutionary models (Figure \ref{HRDFig}; \citealt{UGT16}) predict that RU Vul has declined in luminosity since the onset of the thermal pulse. We lack the historical infrared photometry to show this, but the decrease in optical pulsation period suggests that RU Vul may have declined in radius by $\sim$18 per cent (Section \ref{PropBackDiscSect}). Since its temperature has not measurably changed, its luminosity has also decreased, by $\sim$36 per cent (Section \ref{PropBackDiscSect}). While this does not account for changes in stellar structure, it would imply that the dust-formation radius for any given species will have similarly contracted by $\sim$18 per cent, allowing dust to rapidly condense over a shell of this width. For our chosen dust chemistry, with $R_{\rm in} \approx 13$ R$_\ast$ (Section \ref{DustDriveSect}), that implies that dust is in the process of condensing over a shell $\sim$2.3 R$_\ast$ ($\sim$1.6 AU) in width. Many models of dust formation \citep[e.g.][]{BHAE15} include an oscillating, quasi-stationary layer of dense material around the star, which occupies the inner edge of the dust-formation zone (see also, e.g., \citealt{KWdK+15}). \citet{MBG+19} hypothesised that this layer may normally be denser around metal-poor stars, as it takes longer for grains to grow sufficiently to overcome stellar gravity. If this is the case, dust formation could rapidly occur in this quasi-stationary layer, rapidly increasing the infrared emission as observed. In the model of \citet{BH12}, where dust chemistry changes and dust opacity increases with radius, opaque dust species (e.g., iron-rich silicates and metallic iron) could condense further out in the wind, triggering a rapid increase in dust opacity as well as dust mass.

While the inner parts of the wind will receive additional radiation pressure per unit mass due to dust condensation, this will not be true in the wind's outer regions. Beyond $\sim$60 R$_\ast$, we expect even the most opaque dust species can condense (Section \ref{DustDriveSect}). However (assuming $M > 0.53$ M$_\odot$), a 3 km s$^{-1}$ wind will not achieve escape velocity until 100 AU ($\sim$150 R$_\ast$). In these regions, the decreased luminosity of the star and increased self-shielding by dust close to the star will decrease the radiation pressure available. Consequently, it is possible that these regions will stop expanding entirely and have a net inflow back towards the star until the thermal energy from the thermal pulse reaches the stellar surface in a few centuries' time.

\subsection{A clumpy wind for RU Vul and other metal-poor stars?}
\label{Disc2Sect}

Since the dust does not redden the star in our line-of-sight, unless we invoke a circumstellar/circumbinary disc (Section \ref{DiscBinarySect}), we must instead invoke a clumpy wind structure where there are no clumps in front of the star. Globular cluster stars may show a similar lack of absorption of optical light by circumstellar dust, as their photometric and spectroscopic temperature largely agree \citep[e.g.][]{MBvLZ11,LNH+14}, with only some stars showing significant reddening, potentially caused by clumps in the wind (e.g., 47 Tuc V3, \citealt{MBvLZ11,MBG+19}; $\omega$ Cen V6, \citealt{MvLD+09}).

Clumps may be a natural consequence of marginal winds where dust formation is slow or inefficient, as UV radiation selectively removes dust-precursor molecules in underdense regions, allowing dust to form only in over-dense regions \citep{vdSDL+18}. \citet{TKTY17} hypothesised that these over-dense clumps will reach the critical opacity needed to escape before the surrounding wind, so will convect outwards in the wind. \citet{MBG+19} discuss this in the context of metal-poor stars.

The same effect could be present in RU Vul: a quasi-stationary dust-forming layer accumulates clumps of dust until they are able to gain enough momentum to leave the star. The resulting change in wind structure would exacerbate the unusual $I_{CO(3\rightarrow2)}/I_{CO(2\rightarrow1)}$ ratio by concentrating matter closer to the star, where the CO $J = 3\rightarrow2$ line is stronger. \citet{MBG+19} find a similar high $I_{CO(3\rightarrow2)}/I_{CO(2\rightarrow1)}$ ratio in the globular cluster star 47 Tuc V3, and while here it fits models whereby the wind is dissociated by intra-cluster UV radiation, it could also receive a contribution from such an altered wind structure. Together, the observations of RU Vul and globular cluster stars suggest that the CO (3$\rightarrow$2) line may typically be much stronger in metal-poor stars than naively predicted. This would be part of a more general trend toward higher $I_{CO(3\rightarrow2)}/I_{CO(2\rightarrow1)}$ ratios in optically thin shells \citep[e.g.][]{Olofsson08,RSOL08,DBDdK+10}, and suggests CO (3$\rightarrow$2) represents a much better observational diagnostic of metal-poor stellar winds than CO (2$\rightarrow$1).

\subsection{RU Vul in a class of period-changing stars}
\label{DiscPChangeSect}

\begin{figure}
\centerline{\includegraphics[height=0.45\textwidth,angle=-90]{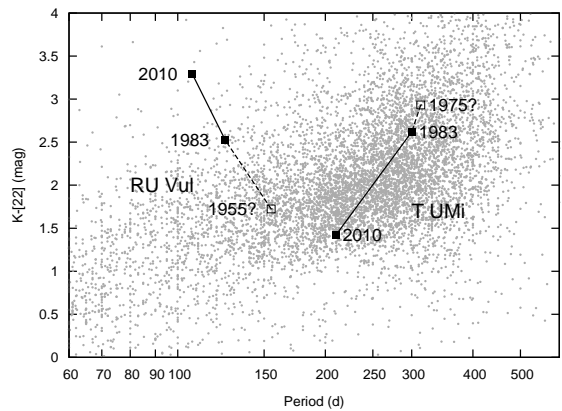}}
\caption{Motion of RU Vul and T UMi in the period--infrared-excess diagram. The background points show the data from the General Catalogue of Variable Stars \citep{SD+04}. The calculation of the solid points are discussed in the text. The extrapolation back to the pre-thermal-pulse period (1955 and 1975, respectively) is done on the assumption that $K-[22]$ has been changing linearly over time.}
\label{PXSFig}
\end{figure}

\citet{MJK19} recently reported on T UMi, which began a phase of period shortening during the 1970s, likely linked to a thermal pulse. Table \ref{IRTable} shows the corresponding infrared observations for T UMi, alongside RU Vul. While RU Vul displays a factor of $\sim$2 increase in infrared flux, T UMi has shown a decrease in infrared flux between the \emph{IRAS} and \emph{Akari}/\emph{WISE} epochs by a factor of $\sim$3. T UMi is well-resolved and clear of any background emission, even in the low-resolution \emph{IRAS} images, and these changes are largely independent of the wavelength observed. This strongly suggests that the cause of infrared dimming in T UMi is related to the destruction of dust around the star.

Figure \ref{PXSFig} shows the motion of RU Vul and T UMi in the period--infrared-excess diagram. To construct this, we have used the historical infrared photometry in Table \ref{IRTable}. The literature data near $K$-band (2.2 $\mu$m) does not allow us to differentiate whether the stars are brightening or fading, so we assume a constant value throughout, i.e., the 2MASS $K_{\rm s}$-band flux for RU Vul and the more-accurate \emph{COBE/DIRBE} flux for T UMi. We construct a proxy for the \emph{WISE} [22] flux in 1983 by interpolating in magnitude and wavelength between the \emph{IRAS} [12] and [25] photometry. While not exact, the uncertainties this imparts ($\sim$0.1 mag) are small compared to the observed changes. To estimate the $K$-[22] colours of both stars before the onset of period change, we can linearly extrapolate backwards in time, using the rate of change of $K$-[22] magnitude between 1983 and 2010. The periods for these epochs come from \citet{UGT16} and \citet{MJK19}, respectively.

Using this method, we can see that both stars appear to start their evolution on the same period--infrared-excess sequence as most dusty stars, but that they take very different tracks through the diagram, with RU Vul becoming more dusty and T UMi becoming less dusty. This indicates a variety of observational outcomes can be expected from the initial phases of thermal pulses, meaning multi-wavelength monitoring of them is important to understand what is occurring with their dust production and destruction.


\section{Conclusions}
\label{ConcSect}

We have made the first detection of a CO envelope around a truly metal-poor evolved star, the pulsating AGB star RU Vul, and report new observations of its infrared spectrum. The star appears to be currently undergoing the early stages of a thermal pulse (Section \ref{PropBackDiscSect}; \citealt{UGT16}): the pulsation period is shortening, and the star is becoming brighter in the optical and mid-infrared. We have modelled the fundamental properties of the star.

We make a clear detection of the CO (3$\rightarrow$2) line, which has a narrow profile, with a half-width at half-maximum of 1.8 km s$^{-1}$ and a half-width at zero intensity of 3.5 km s$^{-1}$. This likely represents an outflow at a speed between these velocities, though we cannot rule out turbulent motion or an inflow. SED modelling shows that the star is experiencing very little circumstellar absorption, so this likely comes from a relative absence of circumstellar material in our line of sight. If RU Vul is a single star, we advocate a clumpy wind structure to explain this, and suggest this may be typical of winds from metal-poor AGB stars. If it has an unseen binary companion, we suggest a face-on circumbinary disc of material.

The mid-infrared spectrum shows a very weak, broad silicate feature. Some emission from Al$_2$O$_3$ may also be present. However, the dominant emission comes from a featureless infrared continuum. This is seen in other metal-poor stars, where the continuum flux is attributed to an unusual kind of dust, possibly metallic iron. This dust may mask features of less opaque dust, including the silicates.

The continuum detections by ALMA at $\sim$1 mm indicate that there is little dust beyond a few hundred R$_\ast$ of the star. Simultaneously, the star is rapidly becoming brighter in the mid-infrared, almost certainly as a result of rapid dust formation close to the star. While this could indicate a spike in mass-loss rate at the start of the current period of change (circa 1965), we suggest it is a consequence of rapid, clumpy dust formation that is occurring as the star fades, following the extinguishment of hydrogen burning by the thermal pulse. We also show that T UMi, which is in similar evolution stage, is evolving down the opposite pathway, with its dust being destroyed.

We advocate further sub-mm observations of metal-poor AGB stars in the Galactic halo to determine how representative RU Vul is of stars of this type, higher-resolution observations of RU Vul (including very-long-baseline maser observations) to probe the kinematics in the wind of this unusual star, and multi-wavelength monitoring of RU Vul and similar stars to determine how their bolometric luminosity and dust production are changing over time.


\section*{Acknowledgements}

The authors acknowledge support from the UK Science and Technology Facility Council under grant ST/L000768/1. This paper makes use of the following ALMA data: ADS/JAO.ALMA\#2016.1.00079.S. ALMA is a partnership of ESO (representing its member states), NSF (USA) and NINS (Japan), together with NRC (Canada), MOST and ASIAA (Taiwan), and KASI (Republic of Korea), in cooperation with the Republic of Chile. The Joint ALMA Observatory is operated by ESO, AUI/NRAO and NAOJ. Based on observations collected at the European Organisation for Astronomical Research in the Southern Hemisphere under ESO programme(s) 099.D-0201(A).





\bsp	
\label{lastpage}
\end{document}